\documentclass[12pt,preprint,showpacs,preprintnumbers,nofootinbib]{revtex4}

\usepackage{graphicx}
\usepackage{epsfig}
\usepackage{dcolumn}
\usepackage{bm}
\usepackage{amssymb}
\usepackage{epsfig}
\usepackage{here}

\renewcommand{\thefootnote}{\fnsymbol{footnote}}

\newcommand{\tx}{\texttt}
\def\beq{\begin{equation}}
\def\eeq{\end{equation}}
\def\bea{\begin{array}}
\def\eea{\end{array}}
\def\be{\begin{equation}}
\def\ee{\end{equation}}
\def\ba{\begin{eqnarray}}
\def\ea{\end{eqnarray}}

\def\ifb{ {\rm fb}^{-1} }

\def\to{\rightarrow}

\def\dis{\displaystyle}
\def\f{\frac}
\def\[{\left[}
\def\]{\right]}
\def\({\left(}
\def\){\right)}

\def\Ah{{\widehat{A}}}
\def\ms{{\widetilde{m}}}
\def\sm0{{\widetilde{m}_0}}
\def\sB{{\sin\beta}}

\def\gs{\tilde{g}}
\def\ts{\tilde{t}}
\def\cs{\tilde{c}}

\def\ov{\overline}

\def\U1em{{U(1)_{\rm em}}}
\def\to{\rightarrow}

\def\sq2{\sqrt{2}}

\def\tanb{\tan\beta}
\def\cotB{\cot\beta}
\def\ee{e^+e^-}

\def\End{\end{document}}

\def\Journal#1#2#3#4{{#1} {\bf #2}, #3 (#4)}

\def\NPB{{\rm Nucl. Phys.} B}
\def\PLB{{\rm Phys. Lett.}  B}
\def\PRL{\rm Phys. Rev. Lett.}

\def\PRD{{\rm Phys. Rev.} D}
\def\ZPC{{\rm Z. Phys.} C}
\def\EPC{{\rm Eur. Phys. J.} C}

\def\thisday{March, 2003}
\begin{document}

\preprint{{\large hep-ph/0209376}}

\title{Single Charged Higgs Boson Production in\\[-1.5mm]
    Polarized Photon Collision and the Probe of New Physics
\vspace*{5mm}
        }%
\author{%
{\sc Hong-Jian He}\,$^1$\footnote{hjhe@physics.utexas.edu},~~
{\sc Shinya Kanemura}\,$^2$\footnote{shinya.kanemura@kek.jp},~~
{\sc C.--P. Yuan\,$^3$}\footnote{yuan@pa.msu.edu}
}
\affiliation{%
\vspace*{5mm}
$^1$Center for Particle Physics,
University of Texas at Austin, Texas 78712, USA\\
$^2$Theory Group, KEK, Tsukuba, Ibaraki 305-0081, Japan\\
$^3$Department of Physics and Astronomy,
Michigan State University, East Lansing, Michigan 48824, USA
\vspace*{30mm}
}

\begin{abstract}
\hspace*{-0.35cm}
We study single charged Higgs boson production
in photon-photon collision as a probe of the new dynamics of
Higgs interactions.
This is particularly important
when the mass ($M_{H^\pm}$) of charged Higgs bosons
($H^{\pm}$) is relatively heavy and above the kinematic limit
of the pair production (\,$M_{H^\pm} > \sqrt{s}/2$\,).
We analyze the cross sections of single charged Higgs boson production
from the photon-photon fusion processes,
$\gamma\gamma\to \tau^- \bar \nu H^+$ and
$\gamma\gamma\to  b   \bar c H^+$, as
motivated by the minimal supersymmetric
standard model and the dynamical Top-color model.
We find that the cross sections at such a $\gamma\gamma$ collider
can be sufficiently large even for
\,$M_{H^\pm} > \sqrt{s}/2$\,, and is typically one to
two orders of magnitude higher than that at its parent
$e^-e^+$ collider.
We further demonstrate that the polarized photon beams can
provide an important means to determine the chirality structure of
Higgs Yukawa interactions with the fermions.
\pacs{\,12.60.-i,\,12.15.-y,\,11.15.Ex
\hfill   ~~ [ \thisday\, ] }

\end{abstract}

\maketitle

\setcounter{footnote}{0}
\renewcommand{\thefootnote}{\arabic{footnote}}


\newpage
\section{Introduction}

The Standard Model (SM) of particle physics demands a single
neutral physical Higgs scalar ($h^0$)\,\cite{Higgs} to generate masses
for all observed weak gauge bosons, quarks and leptons,
while leaving the mass of Higgs boson 
and all its Yukawa couplings unpredicted.
A charged Higgs boson ($H^\pm$) is an unambiguous
signature of the new physics beyond the SM.
Most extensions of the SM require an extended electroweak symmetry
breaking (EWSB) sector with charged Higgs scalars as part of its
physical spectrum at the weak scale.
The electroweak gauge interactions of $H^\pm$
are universally determined by its electric charge and weak-isospin,
while the Yukawa couplings of $H^\pm$ are model-dependent
and can initiate new production mechanisms for $H^\pm$ at high energy
colliders. Most of the underlying theories that describe the EWSB
mechanism can be categorized as either a ``supersymmetric'' (with
fundamental Higgs scalars) \cite{SUSY} or a ``dynamical'' (with
composite Higgs scalars) \cite{DSB} model.
The minimal supersymmetric SM (MSSM) \cite{MSSM} and the dynamical
Top-color model \cite{Hill} are two typical examples.
As we will show, the Yukawa couplings
associated with the third family quarks and leptons can be large
and distinguishable in these models, so that measuring the single
charged scalar production rate in the polarized photon collisions
can discriminate these models of flavor symmetry breaking.

If a charged Higgs boson could be sufficiently light,
with mass ($M_{H^\pm}$) below $\sim 170$\,GeV, it may
be produced from the top quark decay,
$\, t \to H^+ b$~\cite{tbH},  at
the hadron colliders, including the Fermilab Tevatron and the
CERN Large Hadron Collider (LHC).
For $\,M_{H^\pm} > m_t-m_b$,\,
$H^\pm$ can be searched at the Tevatron and the LHC
from the production processes
$g b \to H^- t$ \cite{gbHt},
$c \bar s, c\bar b \to H^+$ \cite{hy,bhy,dhy,xx}, and
$gg$, $q\bar q$
$\to H^\pm W^\mp$\cite{ppHW,ppHW2}, etc.
The associate
production of $H^\pm t$ from $gb$ fusion is difficult to detect
at the Tevatron because of its small rate (largely
suppressed by the final state phase space),
but it should be
observable at the LHC for $M_{H^\pm} \lesssim 1$\,TeV~\cite{gbHt}.
The single $H^\pm$ production from $cs$ or $cb$
fusions is kinematically advantageous so that it can yield a sizable
signal rate, and can be detected at colliders as long as
the relevant Yukawa couplings are not too small\,\cite{hy,dhy}.
The $gg \to H^\pm W^\mp$ process originates from loop corrections,
and is generally small for producing a heavy $H^\pm$
unless its rate is enhanced by $s$-channel resonances,
such as $gg \to H^0 ({\rm or~} A^0) \to H^\pm W^\mp$.\,
Similarly, the rate of $q\bar q \to H^\pm W^\mp$ is small
in a general two-Higgs-doublet model (THDM).
This is because for light quarks in the initial state, this process
can only occur at loop level, and for heavy quarks in the initial state,
this process can take place at tree level via Yukawa couplings but is
suppressed by small parton luminosities of
heavy quarks inside the proton (or anti-proton).
If $H^\pm$ is in a triplet representation, the
$Z$-$H^\pm$-$W^\mp$ vertex can arise from a custodial breaking term
in the tree level Lagrangian, but its strength has to be
small due to the strong experimental constraint on the $\rho$-parameter.
Hence, the production rate of $q\bar q \to Z \to H^\pm W^\mp$
cannot be large either.
At hadron colliders, charged Higgs bosons can also be produced in
pairs via the $s$-channel
$q \bar q$ fusion process through the gauge interactions of
$\gamma$-$H^+$-$H^-$ and $Z$-$H^+$-$H^-$\,
and the $s$-channel gluon fusion process\cite{ppHpHm}.
However, the rate of
the pair production generally is much smaller than that predicted by
the single charged Higgs boson production mechanisms
when the mass of the charged Higgs boson increases.

If $M_{H^\pm}$ is smaller than half of the center-of-mass
energy ($\sqrt{s}$) of a Linear Collider (LC), then $H^\pm$ may be
copiously produced in pairs via the scattering processes
$e^-e^+ \to H^- H^+$ and $\gamma \gamma \to H^- H^+$~\cite{eeHpHm,eeHpHm2}.
The production rate of a $H^- H^+$ pair is determined by
the electroweak gauge interactions of $H^\pm$, which
depends only on the electric charge and weak-isospin of $H^\pm$.
When $\,M_{H^\pm} > \sqrt{s}/2$,\,
it is no longer possible to produce the
charged Higgs bosons in pairs. In this case, the predominant production
mechanism of the charged Higgs boson is via the
single charged Higgs boson production processes, such as
the loop induced process
$e^-e^+ \to H^\pm W^\mp$~\cite{eeWH,KMO1},
and the tree level processes
$e^-e^+ \to b \bar c H^+, \, \tau^- \bar \nu H^+$  and
$\gamma\gamma \to b \bar c H^+, \, \tau^- \bar \nu H^+$\,\cite{hky}.
The production rate of the above tree level processes depends on the
Yukawa couplings of fermions with $H^\pm$.
This makes it possible to discriminate models of flavor
symmetry breaking by measuring the production rate of the single
charged Higgs boson at LC.
However, as to be discussed below, at $e^+e^-$ colliders,
the cross sections of the single $H^\pm$ production processes
induced by the Yukawa couplings of fermions with $H^\pm$
are generally small because single $H^\pm$ events are
produced via $s$-channel processes
(with a virtual photon or $Z$ propagator).
On the other hand, at $\gamma\gamma$ colliders \cite{pc,ecfa,JLC},
the single $H^\pm$ cross sections
are enhanced by the presence of the $t$-channel
diagrams which contain collinear poles in high energy collisions.

It is well known that one of the main motivations for a
high-energy polarized photon collider is
to determine the \tx{CP} property of
the neutral Higgs bosons\,\cite{CPprop1,CPprop2,RevLC,CPprop3}.
In this work, we provide another motivation
for having a polarized photon collider --
to determine the chirality structure of the fermion Yukawa couplings
with the charged Higgs boson via single charged Higgs
boson production so as to discriminate the dynamics of flavor symmetry
breaking.
Specifically, we study single charged Higgs boson production
associated with a fermion pair ($\bar{f'} f$) at photon colliders, i.e.,
$\gamma\gamma\to \bar f' f H^\pm$,
($\bar{f'} f = bc,~{\rm or},~\tau\nu $), based on our recent proposal
in Ref.\,\cite{hky}.
Two general classes of models will be discussed to predict the signal
event rates -- one is the weakly interacting models
represented by the MSSM\,\cite{MSSM}
and another is the dynamical symmetry breaking models
represented by the Top-color (TopC) model\,\cite{Hill}.
We show that the yield of a heavy charged Higgs boson
at a $\gamma\gamma$ collider is typically one to two orders
of magnitude larger than that at an  $e^-e^+$ collider.
Furthermore, we demonstrate that
a polarized photon collider can either enhance or suppress
the single charged Higgs boson production, depending on the
chirality structure of the corresponding Yukawa couplings.

To clarify the physics implication of a polarized photon collider, 
we shall consider in this paper the center-of-mass (c.m.) energy of
a \,$\gamma\gamma$\, collider to be about 80\% of
an \,$e^-e^+$\, collider, and leave a more realistic analysis, that 
takes into account the dependence of the $\gamma \gamma$ luminosity 
and energy on the polarization of the photon beam, to a future
publication.\footnote{
Currently, we are collaborating with the experimentalists, who are
interested in the photon collider option of the Linear Collider Working
Group \cite{lcwg}, to perform a study including detector simulation and
effects due to the energy dependence of the luminosity of the 
polarized photons 
produced from Compton back-scattering.
} 
This approximation is motivated by the fact that
the mean energy of a typical energy spectrum of high energy photons
generated by the Compton back-scattering of a few MeV laser beam is
$E_\gamma \simeq 0.8 E_{e^\pm}$, where $E_{e^\pm}$ is the energy of
the $e^-$ or $e^+$ beam~\cite{telnov}.
Though the detailed distributions of the luminosity and energy of 
the polarized photon beam are strongly model dependent, the gross
feature of those distributions can be studied from a model proposed in 
Ref. \cite{pc}. We show that after 
including the reduction factor for choosing a specific polarization
state of the photon beam, the above approximation agrees within 
a factor of 2 with the 
calculation convoluting the constituent $\gamma \gamma$ cross section
with the luminosity distribution of the polarized photon beam, 
when the dominant
polarization state of the photon beam is considered.
This observation is supported by a calculation presented in
Ref.~\cite{ko} in the context of considering an $e^-\gamma$ process.
The above approximation was found to be in good agreement with
that obtained by folding the constituent cross section with
the luminosity function of the initial state photon~\cite{kmo_eg}.
As to the resolution power of the polarized photon collider, a
convoluted calculation gives a stronger resolution at the cost of
a smaller cross section, which will also be illustrated in Sec. IV.

The rest of the paper is organized as follows.
In Sec. II, we discuss the relevant Yukawa interactions in the weakly
interacting MSSM and the strongly interacting TopC model. 
The production cross section of the single charged Higgs boson in a
polarized collider is presented in Sec. III, which also contains
discussions on how to discriminate MSSM from TopC using a polarized
photon collider. 
Sec. IV contains discussions on the effect of including a model of the 
energy dependent luminosity of the polarized photon beam, as well as
our conclusions.

\section{Yukawa Interactions in MSSM and Top-color Model
}

For generality, we define the charged Higgs Yukawa interaction as
\beq
\label{eq:yukawa}
{\cal L}_{\rm Y} = \ov{f'}
\left( Y_L^{f'f} P_{L} + Y_R^{f'f} P_{R} \right)
       f \,H^- + {\rm h.c.}\,,
\eeq
where $f$ and $f'$ represent up-type and down-type fermions, respectively,
and
$P_{L,R}$ are the chirality projection operators
$ P_{L,R}= \(1\mp \gamma_5\)/2$\,.

We first consider the Yukawa sector of the MSSM, which is similar to
that of a Type-II THDM.
The corresponding tree-level Yukawa couplings of fermions
with $H^\pm$ are given by
\begin{eqnarray}
\label{eq:yukawaC}
Y_{L(0)}^{f'f} \!=\! \frac{\sqrt{2} m_{f'}}{v} V_{ff'}  \tan\!\beta, \;\;\;
Y_{R(0)}^{f'f} \!=\! \frac{\sqrt{2} m_f}{v} V_{ff'}  \cot\!\beta,
\end{eqnarray}
where $m_{f}$ ($m_{f'}$) is  the mass of the fermion
$f$ ($f'$),
\,$\tan\beta = \langle H_u\rangle / \langle H_d\rangle$\,
is the ratio of the vacuum expectation values
(${\langle H_u\rangle}$ and ${\langle H_d\rangle}$)
of the two Higgs doublets with
$v = \sqrt{{\langle H_u\rangle}^2 + {\langle H_d\rangle}^2}
   \simeq 246$\,GeV,
and $V_{ff'}$ is the relevant
Cabibbo-Kobayashi-Maskawa (CKM) matrix element
of the fermions $f$ and $f'$.
The coupling constants
$Y_{L(0)}^{f'f}$ and $Y_{R(0)}^{f'f}$
vary as the input parameter $\tan\beta$ changes.
For instance, for the $\tau^+$-$\nu$-$H^-$ coupling,
$Y_{L(0)}^{\tau\nu}$ increases as $\tan\beta$ grows, and
reaches about \,$0.20-0.51$\, for \,$\tan\beta=20-50$,\, while
$Y_{R(0)}^{\tau\nu}$ is zero because of the absence of
right-handed Dirac neutrinos in the MSSM.
Without losing generality, we shall choose the following
typical inputs
for our numerical analysis:
\beq
\(Y_{L(0)}^{\tau\nu},\, Y_{R(0)}^{\tau\nu}\)
\,=\, (0.3,~0)\,,
~~~~{\rm for}~ \tanb = 30\,.
\label{eq:taunuH-MSSM-def}
\eeq

\begin{figure}[H]
\vspace*{-12mm}
\begin{center}
\hspace*{-3mm}
\includegraphics[width=14cm,height=18cm]{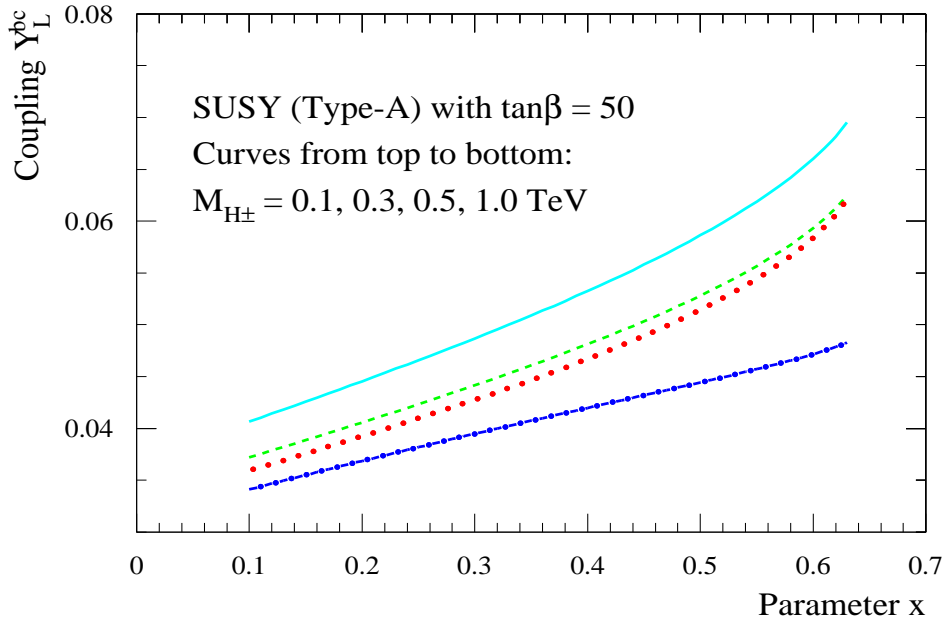}
\end{center}
\vspace*{-8.5cm}
\caption{
The radiative $\bar b$-$c$-$H^-$ coupling
as a function of the parameter $x$
in the minimal Type-A SUSY models with $y=0$.
Here, we set $(m_{\tilde{g}},\mu,\ms_0 )= (300,\,300,\,600)$\,GeV,
and $A=-A_b=1.75$\,TeV. This result also includes the QCD running
effect for the Born level Yukawa coupling.
}
\label{fig:YRbcH}
\end{figure}

The tree level $\bar b$-$c$-$H^-$ coupling
contains a CKM suppression factor $V_{cb}\simeq 0.04$, so that
$Y_{L(0)}^{bc}$ is around $0.03$ for $\tan\beta=50$, and
$Y_{R(0)}^{bc}$ is less than about $2\times 10^{-4}$ for $\tan\beta > 2$.
However, supersymmetry (SUSY) 
radiative corrections can significantly enhance
the tree level $\bar b$-$c$-$H^-$ coupling.
It was shown in Ref.\,\cite{dhy} that
the radiatively generated $\bar b$-$c$-$H^-$ coupling from
the stop-scharm ($\ts-\cs$) mixings
in the SUSY soft-breaking sector
can be quite sizable.
For instance, in the minimal Type-A SUSY models, the
non-diagonal scalar trilinear $A$-term for the
up-type squarks can be written as \cite{dhy}
\beq
A_u ~=~
      \left\lgroup
      \bea{ccc}
      0 ~&~ 0 ~&~ 0\\
      0 ~&~ 0 ~&~ x\\
      0 ~&~ y ~&~ 1
      \eea
      \right\rgroup A \,,
\label{eq:Au}
\eeq
which generates a non-trivial $4\times4$ squark mass-matrix among
$(\cs_L,\,\cs_R,\,\ts_L,\,\ts_R)$.
In $A_u$, the parameters $(x,\,y)$ can be naturally of
order 1,
representing large $\ts-\cs$ mixings that are consistent with
all the known theoretical
and experimental constraints \cite{CCBVS,FCNC}.
An exact diagonalization of
this  $4\times4$ mass-matrix results in the following mass eigenvalues:
\beq
\bea{ll}
M_{\cs1,2}^2 & = \sm0^2 \mp\f{1}{2}|\sqrt{\omega_+}-\sqrt{\omega_-}|\,,
\\[3mm]
M_{\ts1,2}^2 & = \sm0^2 \mp\f{1}{2}|\sqrt{\omega_+}+\sqrt{\omega_-}|\,,
\eea
\label{eq:Mass}
\eeq
with
$
M_{\ts1} < M_{\cs1} < M_{\cs2} < M_{\ts2} \,.
$
Here, $\ms_0$ is a common scalar mass in the diagonal blocks of
the squark mass-matrix,
$~\omega_\pm = X_t^2+(x\Ah\pm y\Ah )^2\,$,
$ X_t = \Ah - \mu\,m_t\,\cotB$ and
$ \Ah = Av\,\sB/\sqrt{2}\,$.
In the squark mass-eigenbasis, the $\bar b$-$c$-$H^-$
coupling can be radiatively induced from the
vertex corrections [scharm(stop)-sbottom-gluino loop]
and the self-energy corrections [scharm(stop)-gluino loop].
In the Type-A models with $ x\neq 0$ and $y=0$,
including the one-loop SUSY-QCD
corrections yields the pattern\,\cite{dhy}:
\beq
 \delta Y_L^{bc}  \neq 0\,  ~~~{\rm and}~~~
 \delta Y_R^{bc}   \simeq  0\,,
\label{eq:A1-cbH}
\eeq
for a moderate to large $\tanb$.
(As to be shown below, this pattern is opposite to that predicted
in the dynamical Top-color model.)
The coupling $ Y_L^{bc}$ is a function of
the mixing parameter $x$,
the Higgs mass $M_{H^\pm}$,
the gluino mass $M_{\gs}$ and the relevant squark masses.
In Fig.~\ref{fig:YRbcH}, we show  $ Y_L^{bc} $
as a function of the parameter $x$
for a typical set of SUSY inputs,
$(m_{\tilde{g}},\,\mu,\,\ms_0 )= (300,\,300,\,600)$\,GeV,
$A=-A_b=1.75$\,TeV, and $\,\tanb=50\,$.
In this figure,
we have also included the QCD running effects for
the tree-level Yukawa couplings, cf. Eq.\,(\ref{eq:yukawaC}) \cite{QCDeff}.
We find that the magnitude of the total coupling $ Y_L^{bc} $ can be
naturally in the range of $0.03-0.07$
for a moderate to large $\tanb$\,.
For a smaller
value of $\tan \beta$, the coupling $ Y_L^{bc} $ decreases.
For instance, for $\tanb=20$, the value of $ Y_L^{bc}$ is about
half of that shown in Fig.~\ref{fig:YRbcH}.

In addition to the SUSY radiative corrections discussed above,
which are not suppressed
by the small CKM matrix element $V_{cb}$,
there are corrections proportional to $V_{cb}$, similar to
those present in the production of $\phi^0 b {\bar b}$
($\phi^0=h^0,H^0,A^0$) with large $\tanb$~\cite{hbb,efflag}.
This effect can be formulated by the corresponding effective
Lagrangian\,\cite{dmb},
\begin{equation}
\mathcal{L\,} ~=~
\frac{\,\sqrt{2}\, V_{cb}\,}{v}\,\frac{
~\overline{m}_{b}(\mu^{~}_{R})\tan\beta~}{1+\Delta_{b}}
\,H^{+}\overline{c_L}
\,b_{R} \,+\, {\rm h.c.} \, ,
\label{eq:efflag}
\end{equation}
where $\mu_{R}$ is the relevant renormalization scale at which we
evaluate the bottom quark running mass $\overline{m}_{b}(\mu_{R})$
including the NLO QCD contributions in the
$\overline{\rm MS}$ scheme\cite{QCDeff}.
In the on-shell scheme, the bare mass of the bottom quark
$m_{bare}$ is equal to $m_{b}+\delta m_{b}$, where $m_b$
is the pole mass and  $\delta m_{b}$ the counter term.
A straightforward calculation shows that the threshold
corrections to $\Delta_{b}$ originating from the SUSY-QCD and
SUSY-electroweak (SUSY-EW) contributions are equal to
$-\delta m_{b}/m_{b}$.
In general,
the SUSY-EW correction comes from loop contributions induced by
the Yukawa and electroweak gauge interactions, where the latter
contribution is usually smaller than the former contribution.
(Since in the generic Type-A model the trilinear term $A$ needs not
to be much smaller than \,$\mu\tanb$,\, we will not make the approximation
\,$A_b -\mu\tanb \approx -\mu\tanb$ \cite{efflag}
in $\Delta_{b}$.)
The SUSY QCD correction is given by
the finite contributions of the sbottom-gluino loop due to
the left-right mixings in the squark-mass matrix\,\cite{dmb},
\begin{eqnarray}
\left( \Delta_{b}\right) _{\mathrm{SUSY-QCD}} & =&
-\,{
\frac{~C_F\alpha_{s}(\mu_R^{~})~}{2\pi}}\,
   m_{\tilde{g}}\,M_{LR}^{b}\,
{\cal I}(m_{\tilde {b}_{1}},\,m_{\tilde{b}_{2}},\,m_{\tilde{g}})
\,,
\label{eq:dmbQCD}
\end{eqnarray}
where
\,$C_{F}=\dis\f{1}{2}\( N_{c}-\f{1}{N_c}\) = \f{4}{3}$\,
with $N_{c}=3$,~
$\alpha_{s}\simeq 0.09$\, at the scale of
$\mu_R^{~}=M_{H^\pm}= O(100)$\,GeV, and
$M_{LR}^{b}=A_{b}-\mu\,\tan\beta$\,.
The SUSY-Yukawa correction to $\Delta_{b}$ arises
from similar loops involving the stop and charged
higgsinos \,$\widetilde{H}_{1,2}$\,, and 
\begin{eqnarray}
\left( \Delta_{b}\right) _{\mathrm{SUSY-Yukawa}}
& =& + \frac{~m_t^2~}{~8\pi^2 v^2~} \f{\mu}{\tanb} M_{LR}^{t}\,
{\cal I}(m_{\tilde {t}_{1}},\,m_{\tilde{t}_{2}},\,\mu)  \, ,
\label{eq:dmbEW}
\end{eqnarray}
where
$M_{LR}^{t}=A_{t}-\mu\cot\beta$.\,
In the above formula, we have defined
\begin{eqnarray}
{\cal I}({m_1},\,{m_2},\,{m_3}) &  =&
-\frac{\dis ~m_{1}^2\,m_{2}^2\ln\frac{m_1^2}{m_2^2}
+m_{2}^2\,m_{3}^2\ln\frac{m_2^2}{m_3^2} + m_3^2\,m_1^2
\ln\frac{m_3^2}{m_1^2} ~}
{(m_1^2-m_2^2)\,(m_2^2-m_3^2)\,(m_3^2-m_1^2)}\, ,
\end{eqnarray}
which, in the special case of
\,$m_1=m_2=m_3\equiv M$,\,
equals to \,$\dis\f{1}{2M^2}$\,.

With the sample values of the SUSY-parameters
given in the caption of Fig.\,\ref{fig:YRbcH},
$\Delta_{b}$ is found to be about $0.17$,
among which, \,$0.20$\, comes from
the SUSY-QCD contribution,  $0.00011$ from
SUSY-Yukawa contribution, and \,$-0.022$\, from
the electroweak gauge contribution\footnote{
The electroweak gauge contribution depends also on other
SUSY parameters \cite{dmb}. Here, $M_2$ is taken to be
$300$\,GeV, but higher values of $M_2$ will make the electroweak
gauge contribution even smaller due to the decoupling feature
of the MSSM.
}.
Hence,  $\Delta_{b}$ yields a factor
of $1/(1+0.17)\simeq 0.85$ suppression in the $b$-$c$-$H^+$ coupling
as compared to the
QCD-improved Born level coupling
(which is about $0.03$ for a 300\,GeV
charged Higgs boson),
and the coupling of $H^+$-$\overline{c_L}$-$b_R$ in
Eq.\,(\ref{eq:efflag})
is about $0.026$ for this set of SUSY parameters.
In other words, the threshold correction
due to the SUSY-QCD and SUSY-EW contributions to $Y_L^{bc}$ is
$(0.85-1)\times 0.03\simeq -0.0045$, which
is not significant in the current case.
(When the SUSY parameter $\mu$ or $A$
flips sign while holding the other parameters
fixed, the threshold correction
from $1/(1+\Delta_b)$ becomes an enhancement rather than
suppression factor.)
The additional contribution to $Y_L^{bc}$ arising from the
$\ts-\cs$  mixing can be read out from Fig.\,\ref{fig:YRbcH} after
subtracting the strength of the QCD-improved Born level coupling.
For instance, using the same set of SUSY parameters described above,
the radiative correction from
$\ts-\cs$ mixings with $x=0.44$ enhances the $ Y_L^{bc}$ coupling by
an amount of $0.02$ ($= 0.05-0.03$) for $M_{H^\pm}=300$\,GeV.
Therefore, the coupling of $Y_L^{bc}$, after including the
QCD-improved Born level coupling $(0.03)$, the radiative correction from
$\ts-\cs$ mixings $(0.02)$, and the threshold correction
due to the SUSY-QCD and SUSY-EW contributions $(-0.0045)$, is about
\,$0.046\,(\simeq 0.05)$\, for the sample SUSY parameters we have chosen.
Hence, without losing generality,
in the following numerical analysis, we choose\footnote{
This choice of couplings may also be realized for lower
$\tan\!\beta$ region with the proper choice of the parameter
$x$ accordingly, {\it e.g.,} for $\tan\!\beta =30$, 
Eq.\,(\ref{eq:bcH-MSSM-def}) corresponds to 
$x\approx 0.6$.  
}
\beq
( Y_L^{bc},\, Y_R^{bc}) = (0.05,\,0) \,
\label{eq:bcH-MSSM-def}
\eeq
as the sample couplings for the MSSM with natural $\ts-\cs$ mixings,
which correspond to the Type-A SUSY models with
$x=O(1)$ and $y=0$ as defined in Ref.\,\cite{dhy}.
(The total decay width of $H^\pm$ will be evaluated for $\tanb=50$.)
It is worth to mention that
the sample flavor-changing $b$-$c$-$H^\pm$ coupling
(\ref{eq:bcH-MSSM-def}) is about a factor-6 smaller than the
sample $\tau$-$\nu$-$H^\pm$ tree-level coupling
(\ref{eq:taunuH-MSSM-def}).

We then consider the dynamical Top-color model \cite{Hill},
which is strongly motivated by the experimental fact
that the observed large top quark mass
($m_t\simeq \dis\f{v}{\sqrt{2}}\simeq 174$\,GeV)
is right at the weak scale,
distinguishing the top quark from all other SM fermions.
This scenario explains the
top quark mass from the $\langle \bar{t}t \rangle$ condensation
via the strong $SU(3)_{\rm tc}$ TopC interaction at the TeV scale.
The associated strong tilting $U(1)$ force is attractive in the
$\langle \bar{t}t \rangle$ channel and repulsive in the
$\langle \bar{b}b \rangle$ channel, so that the bottom quark
mainly acquires its mass from the TopC instanton contribution
\cite{Hill}.
This model predicts three relatively light physical
top-pions $(\pi_t^0,\,\pi^\pm )$.
The Yukawa interactions of these top-pions
with the third family quarks are given by the Lagrangian,
\beq
\bea{l}
\dis\f{m_t\tanb}{v}\hspace*{-1.1mm}\left[
i{K_{UR}^{tt}}
{K_{UL}^{tt}}^{\hspace*{-1.3mm}\ast}\overline{t_L}t_R\pi_t^0
\hspace*{-0.7mm}+\hspace*{-0.8mm}\sq2
{K_{UR}^{tt}}{K_{DL}^{bb}}^{\hspace*{-1.3mm}\ast}
\overline{b_L}t_R\pi_t^- +
 \right.
\\[3.3mm]
\hspace*{1.3cm}~~\left.
i{K_{UR}^{tc}}
 {K_{UL}^{tt}}^{\hspace*{-1.3mm}\ast}\overline{t_L}c_R\pi_t^0
\hspace*{-0.7mm}+\hspace*{-0.8mm}\sq2
{K_{UR}^{tc}} {K_{DL}^{bb}}^{\hspace*{-1.3mm}\ast}
\overline{b_L}c_R\pi_t^-
\hspace*{-0.7mm}+\hspace*{-0.7mm}{\rm h.c.}  \right],
\eea
\label{eq:Ltoppi}
\eeq
where {\small $\tanb = \sqrt{(v/v_t)^2-1}$} and
the top-pion decay constant
$v_t\simeq O(60-100)$~GeV.\footnote{
Note that this $\tan \beta$ does not have the same meaning as the
$\tan \beta$ in the MSSM.
}
The rotation matrices $K_{UL,R}$ and $K_{DL,R}$ are needed
for diagonalizing the up- and down-quark mass matrices
$M_U$ and $M_D$, i.e.,
{\small $~K_{UL}^\dag M_U K_{UR} = M_U^{\rm dia}~$} and
{\small $~K_{DL}^\dag M_D K_{DR} = M_D^{\rm dia}$},~
from which the CKM matrix is defined as
\,{\small $V=K_{UL}^\dag K_{DL}$}\,.\,
As shown in Ref.\,\cite{hy}, to yield a realistic form of
the CKM matrix $V$ (such as
the Wolfenstein-parametrization), the TopC model
generally has the following features:
\beq
\bea{l}
K_{UR}^{tc}\lesssim 0.11\hspace*{-0.5mm}-\hspace*{-0.5mm}0.33~,~
{\rm with}~~~
\\[3.3mm]
K_{UR}^{tt}\simeq 0.99\hspace*{-0.5mm}-\hspace*{-0.5mm}0.94~,~~~
{\rm and}~~~
K_{UL}^{tt} \simeq K_{DL}^{bb} \simeq 1 ~,
\eea
\label{eq:KURtc}
\eeq
which suggests that the $t_R$-$c_R$ transition
can be naturally around $10-30\%$.
Combining Eqs.~(\ref{eq:Ltoppi}) and (\ref{eq:KURtc}),
we can deduce the Yukawa couplings of fermions with
the charged top-pion
(also called charged Higgs boson throughout this paper) as
\beq
\bea{l}
Y_L^{bt} \,=\, Y_L^{bc} \,=\, 0\,,
\\[3.3mm]
Y_R^{bt} \,\simeq \, \dis\f{\sqrt{2}m_t}{v}\tanb \,,~~~~~
Y_R^{bc} \,\simeq \, Y_R^{bt} K_{UR}^{tc} \,.
\eea
\eeq
Thus, taking a typical value of $\tanb $ to be $3$ and
a conservative input for the \,$t_R-c_R$\, mixing
$ K_{UR}^{tc}$ to be $0.1 $ in the TopC model,
we obtain
\beq
Y_R^{bt} \simeq 3\,,
~~~~{\rm and}~~~~
\(Y_L^{bc},\,Y_R^{bc}\)
= (0,\, 0.3) \,,
\label{eq:bcH-TopC-def}
\eeq
which will be used as the sample
TopC parameters for our numerical analysis.
We note that in contrast to the radiative
coupling  of
the charged Higgs boson predicted in the Type-A SUSY model
with $y=0$
(in which $Y_L^{bc} \neq 0 $ and $\, Y_R^{bc}  \simeq  0$, i.e.,
 mainly left-handed),
the charged top-pions only have a right-handed coupling.
This feature of the TopC
 is also opposite to the tree-level $\tau$-$\nu$-$H^\pm$
coupling (which is purely left-handed)
predicted in the MSSM [cf. Eq.\,(\ref{eq:taunuH-MSSM-def})].
As we will demonstrate below, this feature
makes it possible to discriminate the dynamical TopC model from the
MSSM or a Type-II THDM
by measuring the production rates of a single charged Higgs
boson at polarized photon colliders.
Finally, we note that apart form the opposite chirality structures
of the $H^\pm$ Yukawa interactions,  the magnitude of the
sample Top-color $b$-$c$-$H^\pm$ coupling chosen in
(\ref{eq:bcH-TopC-def}) is the same as that of
the sample $\tau$-$\nu$-$H^\pm$ coupling
(\ref{eq:taunuH-MSSM-def}).

\section{
$H^\pm$ Production in ${\gamma}{\gamma}$
Collision as a Probe of New Physics}

\begin{figure}[H]
\begin{center}
\hspace*{-3mm}
\includegraphics[width=11.5cm,height=11cm]{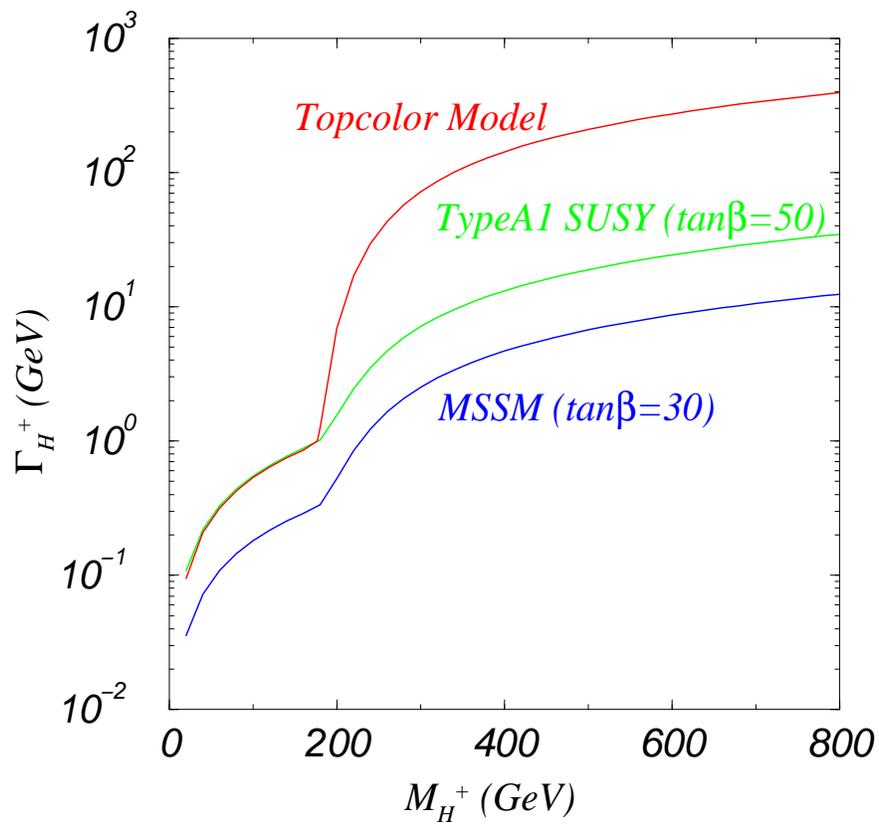}
\end{center}
\vspace*{-5mm}
\caption{The total decay widths of $H^+$
predicted by the models discussed in the text.
}
\label{fig:width}
\end{figure}

\begin{figure}[H]
\vspace*{-8mm}
\begin{center}
\hspace*{-3mm}
\includegraphics[width=11.5cm,height=11cm]{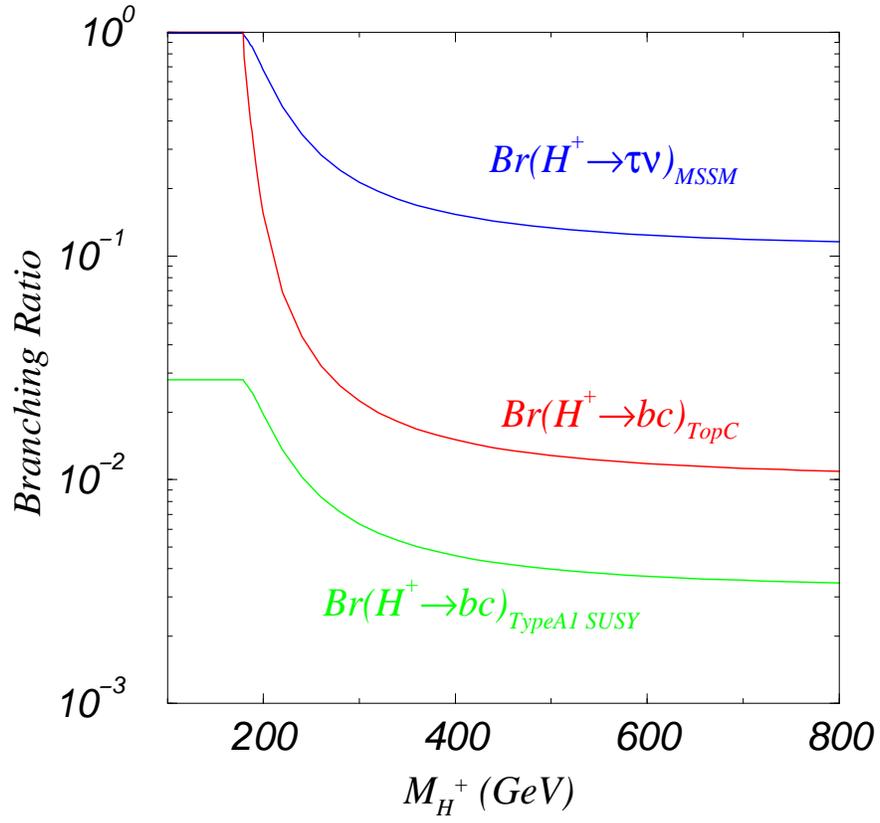}
\end{center}
\vspace*{-5mm}
\caption{The relevant decay branching ratios of $H^+$
predicted by the models discussed in the text.
}
\label{fig:branch}
\end{figure}

We calculate the cross section of
$\gamma\gamma\to \bar f' f H^+$
using the helicity amplitude method for
$f \bar f'=b \bar c \,$  or $\tau^-\bar\nu$.
For the $b \bar c$ channel, we will consider
both the MSSM (with stop-scharm mixings)
and the TopC model using the sample parameters
listed in Eqs.~(\ref{eq:bcH-MSSM-def}) and (\ref{eq:bcH-TopC-def}),
respectively.
For the $\tau^-\bar\nu$ channel, we will
consider the MSSM with the sample parameters given in
Eq.~(\ref{eq:taunuH-MSSM-def}).
The cross sections for other values
of couplings, different from our sample inputs, can
be estimated by a proper rescaling.
In order to predict the event rate
of $\,\gamma\gamma\to \bar f' f H^+$,\, we need to specify the
total decay width $\Gamma_{H^+}$ for $H^\pm$, from which the
decay branching ratio of \,$ H^\pm \to f' f$\, can be calculated.
For simplicity, we shall only include
the quark and lepton decay modes of $H^\pm$ to evaluate
$\Gamma_{H^+}$.
Its bosonic decay modes are not included because
their contributions are generally small and strongly depend on the
other parameters of the model. For example,
in the MSSM, the partial decay width of $H^\pm \to W^\pm h^0$
 depends on the neutral Higgs boson
mixing angle $\alpha$ and the light \tx{CP}-even Higgs boson mass
$m_{h}$, but 
it is generally small, especially when
 $M_{H^\pm}$ becomes large which corresponds to the decoupling limit.
We will also neglect all the loop-induced decay modes such as
$H^\pm \to W^\pm Z$ \cite{hwz},
and assume that the relevant sparticles
are relatively heavy so that the SUSY decay channels of $H^\pm$
are not kinematically accessible.
Finally,
in the TopC model, only the dominant $tb$ and $cb$ decay modes
are included in the calculation of \,$\Gamma_{H^+}$\,.
For the later analysis and discussion,
we show the predicted total decay widths and the relevant
decay branching ratios of $H^+$ in Figs.~\ref{fig:width}
and \ref{fig:branch} as the Higgs mass  $M_{H^\pm}$ varies.

In our numerical analysis, the dominant QCD corrections are included
in the Yukawa couplings by using the running quark masses.
For instance, at the 100\,GeV scale, the running masses of
the bottom and charm quarks
are $m_b=2.9$ GeV and $m_c=0.6$ GeV, respectively.


\begin{figure}[H]
\vspace*{-9mm}
\begin{center}
\includegraphics[width=12cm,height=10cm]{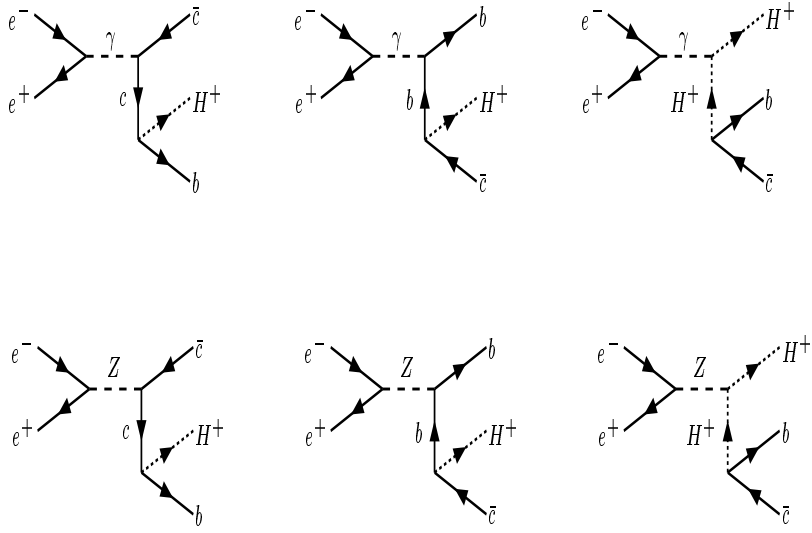}
\end{center}
\vspace*{-12mm}
\caption{
The complete set of Feynman diagrams for $e^-e^+ \to b \bar c H^+$.
}
\label{fig:eefyn}
\end{figure}
\begin{figure}
\begin{center}
\includegraphics[width=16cm,height=16cm]{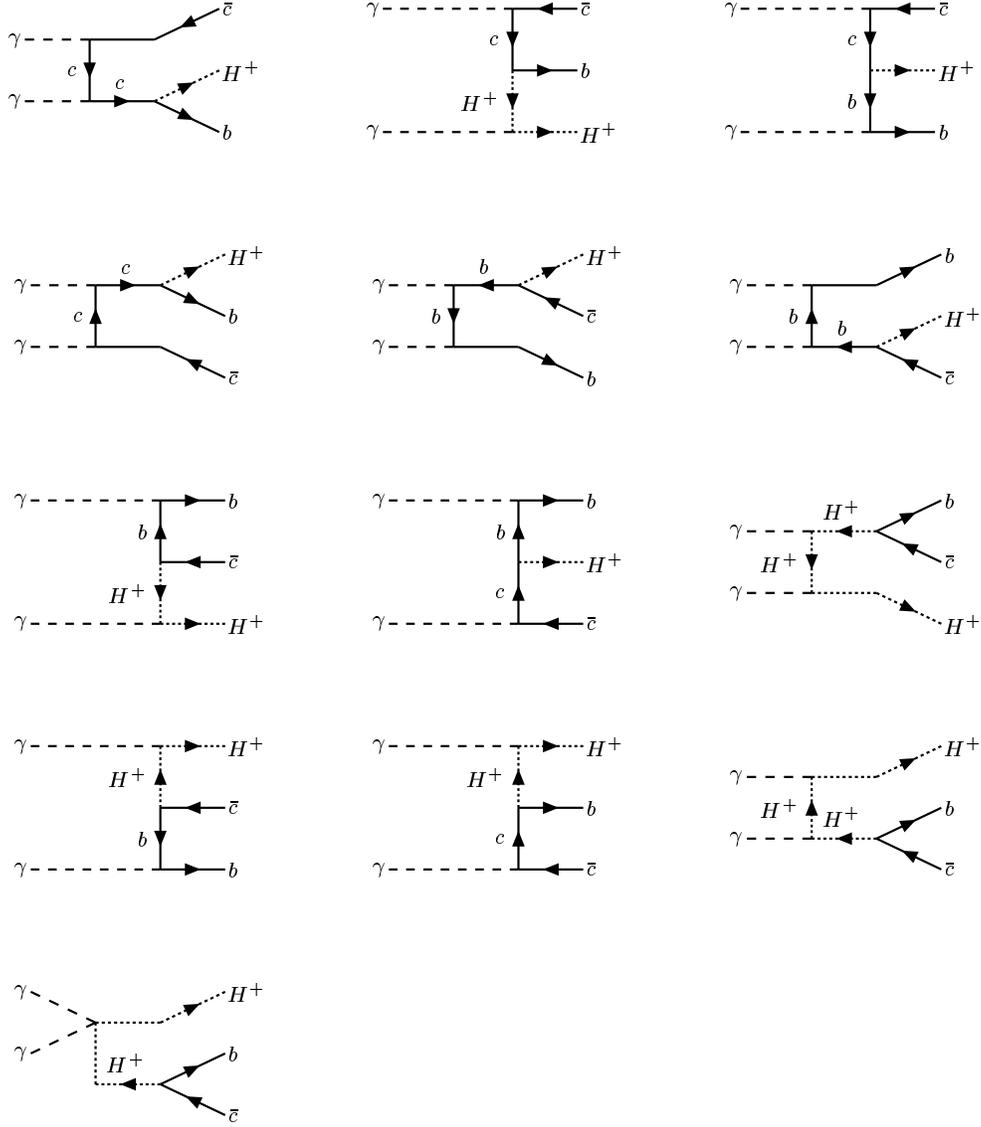}
\end{center}
\vspace*{-5mm}
\caption{
The complete set of Feynman diagrams for $\gamma\gamma \to b \bar c H^+$.
}
\label{fig:aafyn}
\end{figure}

\subsection{$bc H^\pm$ Production}

\begin{figure}[H]
\begin{center}
\includegraphics[width=13cm,height=11cm]{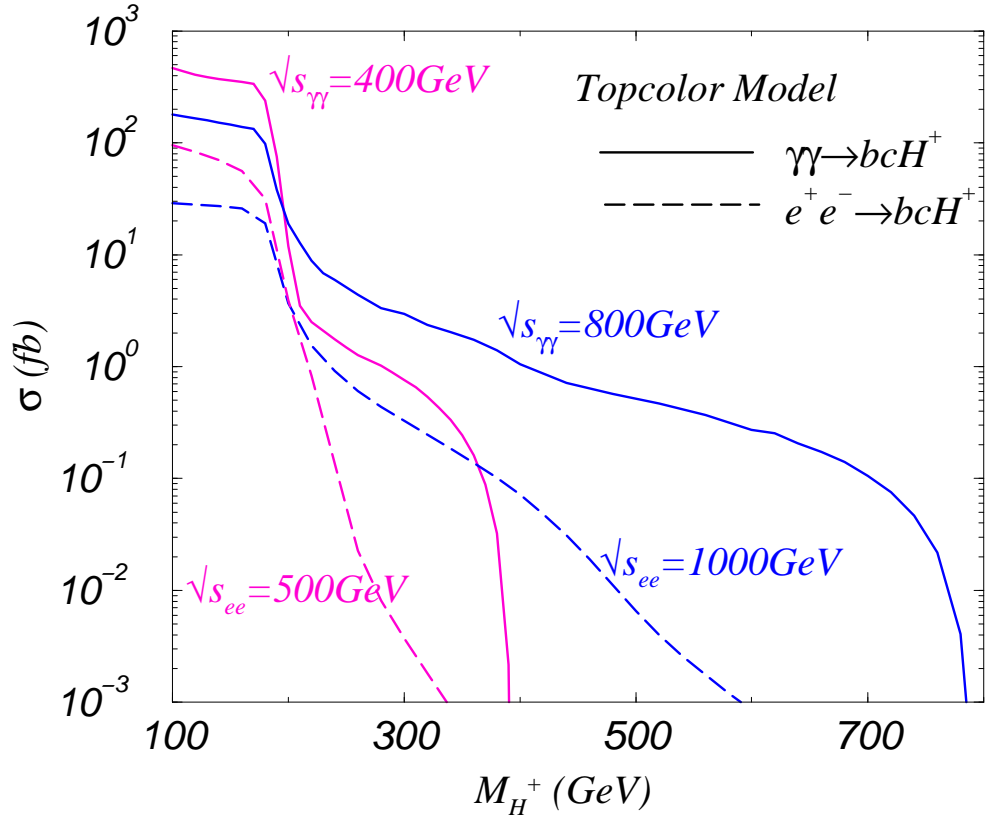}
\end{center}
\caption{Cross sections of $\gamma\gamma\to b \bar c H^+$ (solid curve)
         and $e^+e^-\to b \bar c H^+$ (dashed curve) for the TopC model
[cf. Eq.\,(\ref{eq:bcH-TopC-def})]
         with unpolarized photon beams
         at $\sqrt{s_{\gamma \gamma}}=400$ GeV and $800$ GeV.
}
\label{fig:bch_topc_tot}
\end{figure}

\begin{figure}[H]
\begin{center}
\includegraphics[width=13cm,height=11cm]{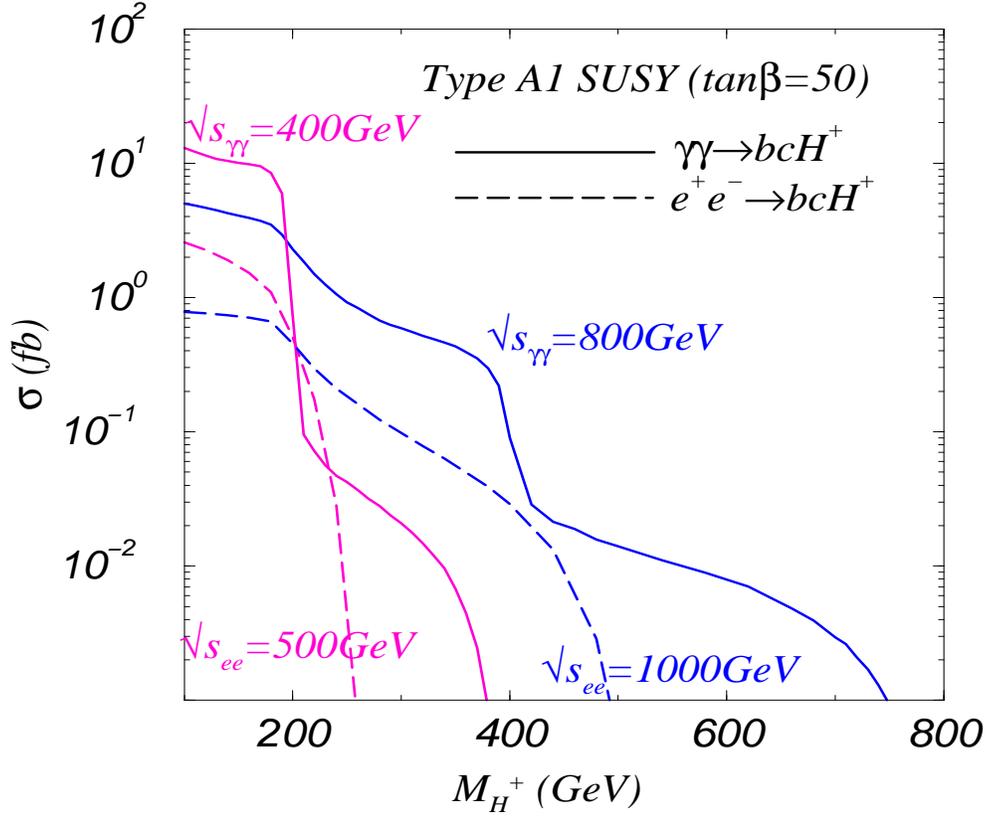}
\end{center}
\caption{Same as Fig.~\ref{fig:bch_topc_tot}, but for the MSSM with
stop-scharm mixings, i.e. Type-A SUSY model
 [cf. Eq.\,(\ref{eq:bcH-MSSM-def})].
}
\label{fig:bch_mssm_tot}
\end{figure}

Using the default parameters of the models as
described in Section\,II, we calculate
the total cross sections of
$e^+e^- \to b \bar c H^+$  and
$\gamma\gamma \to b \bar c H^+$
as a function of $M_{H^\pm}$.
The complete set of Feynman diagrams for the above processes
are depicted in
Figs.\,\ref{fig:eefyn} and \ref{fig:aafyn}, respectively.
The result for the TopC model
is shown in Fig.\,\ref{fig:bch_topc_tot}, where,
for comparison, we have taken the center-of-mass energy 
($\sqrt{s}$) of
the $\gamma\gamma$ collider to be 0.8 times of that of
the $e^-e^+$ collider.
The result for the MSSM with stop-scharm mixings
can be easily obtained from  Fig.\,\ref{fig:bch_topc_tot} by
rescaling the cross sections by a factor
\,$(0.05/0.3)^2=1/36$\, when \,$M_{H^\pm} > \sqrt{s}/2$\,.\,
For \,$M_{H^\pm} < \sqrt{s}/2$,\, where the pair production mechanism
dominates, the actual rate also depends on the
decay branching ratio Br$(H^- \to b {\bar c})$
and the total decay width $\Gamma_{H^+}$ in the MSSM.
For completeness, we also show the result for the MSSM in
Fig.\,\ref{fig:bch_mssm_tot}, which is qualitatively similar to
Fig.\,\ref{fig:bch_topc_tot} except near the boundary of
the available phase space for pair production, i.e. when
$M_{H^\pm} \sim \sqrt{s}/2$.
This is because the total decay width of $H^\pm$
in the TopC model is much larger than that in
the Type-A SUSY mode. For instance, the $\Gamma_{H^+}$
of the charged Higgs boson with a mass $200$\,GeV ($400$\,GeV)
is about $7$\,GeV ($143$\,GeV) in the TopC model
[cf. Eq.\,(\ref{eq:bcH-TopC-def})], and
$1.5$\,GeV ($13$\,GeV) in the Type-A SUSY model
[cf. Eq.\,(\ref{eq:bcH-MSSM-def})].
The branching ratios for the
decay mode \,$H^+ \to c {\bar b}$\, predicted in these two models
 are $\,0.15~ (0.015)$\, and
     $\,0.02~ (0.0046)$,\, respectively.

A few discussions on the feature of the results shown in
Fig.\,\ref{fig:bch_topc_tot} are in order.
(The same discussions also apply to Fig.\,\ref{fig:bch_mssm_tot}.)
For \,$M_{H^\pm} < \sqrt{s}/2$\,,\,
the charged Higgs pair production is kinematically allowed.
In this case, the production cross section of 
$\gamma\gamma \to b \bar c H^+$
( and  $e^-e^+ \to b \bar c H^+$) is dominated by the contribution
from the pair production diagrams with the produced $H^-$ decaying
into a $b \bar c$ pair.
Hence, its rate is proportional to the decay branching ratio
Br$(H^- \to b \bar c)$.
As shown in the figure, there is a {\it kink} structure when
$M_{H^\pm}$ is around 180\,GeV. That is caused by the change in
Br$(H^- \to b \bar {c})$ when the decay channel
 $H^- \to b \bar {t}$ becomes available.
 We also note that the cross section at a higher energy collider,
 either an $e^-e^+$ or $\gamma \gamma$ collider, is larger for 
 the production of a heavy $H^+$ because of the larger  
 final state phase space volume.
  On the other hand, when \,$M_{H^\pm} \ll \sqrt{s}/2$\,,\,
 the cross section approximately scales as $1/{s}$, for the pair production 
process dominates the production rate.
 In Fig.\,\ref{fig:bch_mssm_tot},
the cross section of $\gamma \gamma \to b {\bar c} H^+$
drops around
$M_{H^+} = \sqrt{s}/2$, for
the on-shell $H^-H^+$ pair production mode is closed
when $M_{H^+} > \sqrt{s}/2$.
Moreover, a careful examination reveals that
the cross section of $\gamma \gamma \to b {\bar c} H^+$
drops much more in Fig.\,\ref{fig:bch_mssm_tot} than
in Fig.\,\ref{fig:bch_topc_tot}.
This is because in our calculation we have included the complete
gauge invariant set of Feynman diagrams whose contribution also
depends on the width of the charged Higgs boson.
Since the total decay width of $H^\pm$ in the TopC model is
much larger than that in the Type-A SUSY model
(cf. Fig.\,\ref{fig:width}),
the similar drop in Fig.\,\ref{fig:bch_topc_tot} is much
less noticeable.

It is evident that the cross section of
 \,$\gamma\gamma \to b \bar c H^+$\,
is larger than that of \,$e^+e^- \to b \bar c H^+$\, in the whole
$M_{H^\pm}$ region. For \,$M_{H^\pm} < \sqrt{s}/2$,\, the cross
section in \,$\gamma\gamma$\, collisions is typically 
a factor of 3 to 5  
larger than that in \,$e^-e^+$\, collisions. 
This can be explicitly checked by comparing the
helicity amplitudes of 
$\gamma\gamma\to H^+H^-$ and $e^-e^+ \to H^+H^-$.
The helicity amplitudes for the $H^+H^-$ pair production
in polarized photon collisions are found to be\footnote{
We have checked that our unpolarized cross section
agrees with that in Ref.~\cite{eeHpHm2}.}
\begin{eqnarray}
\label{eq:pair}
M(\gamma_{\lambda_1}^{~} \gamma_{\lambda_2}^{~} \to H^+H^-)
&=&
 - 2 e^2  \lambda_1 \lambda_2
\frac{\xi^2 \/ \sin^2\Theta}{~1 - \xi^2 \cos^2 \Theta~ }
  +  e^2 \left( 1 + \lambda_1 \lambda_2 \right),
\end{eqnarray}
where the degree of polarization of the initial state photons,
$\lambda_{1}$ and $\lambda_{2}$, can take
the value of either $-1$ or $+1$,
corresponding to a left-handedly ($L$) and right-handedly ($R$)
polarized photon beam, respectively;
\,$\Theta$\, is the scattering angle of $H^+$ in the center-of-mass
frame; and  $\, \xi=\sqrt{ 1 - 4 M_{H^\pm}^2 / s }$\,.\,
In the massless limit, i.e., when $M_{H^\pm} \to 0$, 
$\xi \to 1$ and the above result
reduces to
\,$M(\gamma_{\lambda_1}^{~}\gamma_{\lambda_2}^{~}\to H^+H^-) \simeq
 e^2 \left( 1 - \lambda_1 \lambda_2 \right)$\,,
 which yields a flat angular distribution. 
The two non-vanishing helicity amplitudes 
of $e^-e^+ \to H^+ H^-$, for $s \gg m_e^2$, are
\begin{eqnarray}
\label{eq:eepair}
M(e^-_L e^+_R \to H^+H^-)&=&
 - e^2 \xi \sin\Theta \left[ 1 + 
 \frac{(c_w^2-s_w^2)^2}{4 c_w^2 s_w^2} \frac{s}{s-M_Z^2+i M_Z \Gamma_Z } 
 \right],
\nonumber\\[3mm]
M(e^-_R e^+_L \to H^+H^-)&=&
 - e^2 \xi \sin\Theta \left[ 1 - 
 \frac{c_w^2-s_w^2}{2 c_w^2} \frac{s}{s-M_Z^2+i M_Z \Gamma_Z } 
 \right],
\end{eqnarray}
where $e^-_L$ ($e^-_R$) denotes a left-handed (right-handed)
electron; $c_w=\cos\theta_w$ and $s_w=\sin\theta_w$ with $\theta_w$
being the weak mixing angle; and 
$M_Z$ and $\Gamma_Z$ are the mass and width 
of the $Z$ boson, respectively.

For \,$M_{H^\pm} > \sqrt{s}/2$,\, where
the pair production is not kinematically allowed,
 the difference between the cross sections of 
$e^-e^+\to b\bar c H^+$ and $\gamma\gamma\to b \bar c H^+$ 
 becomes much larger
 (two to three orders of magnitude) for a larger
$M_{H^\pm}$ value.
To understand the cause of this difference, we have to examine
the Feynman diagrams, cf.  Figs.\,\ref{fig:eefyn} and \ref{fig:aafyn},
that contribute to the scattering processes
\,$e^-e^+ \to b \bar c H^+$\,
and
\,$\gamma\gamma \to b \bar c H^+$ .\,
In the former process, all the Feynman diagrams contain an $s$-channel
propagator which is either a virtual photon or a virtual $Z$ boson.
Therefore, when $M_{H^\pm}$ increases for a fixed $\sqrt{s}$, the
cross section decreases rapidly.
On the contrary, in the latter process, when $M_{H^\pm} > \sqrt{s}/2$,
the dominant contribution arises from the fusion diagram
$\gamma\gamma \to (c \bar c) (b \bar b)
\to  b \bar c H^+$, whose contribution is enhanced by the
two collinear poles (in a $t$-channel diagram)
generated from  $\gamma \to c \bar c$ and
$\gamma \to b \bar b$ in high energy collisions.
Since the collinear enhancement takes the form of
$\,\ln (M_{H^\pm}/m_q)\,$,
with $m_q$ being the bottom or charm quark mass,
the cross section of $\gamma\gamma \to b \bar c H^+$
does not vary much as $M_{H^\pm}$ increases until it is close
to \,$\sqrt{s}$\,.

\begin{figure}[H]
\begin{center}
\includegraphics[width=13cm,height=11cm]{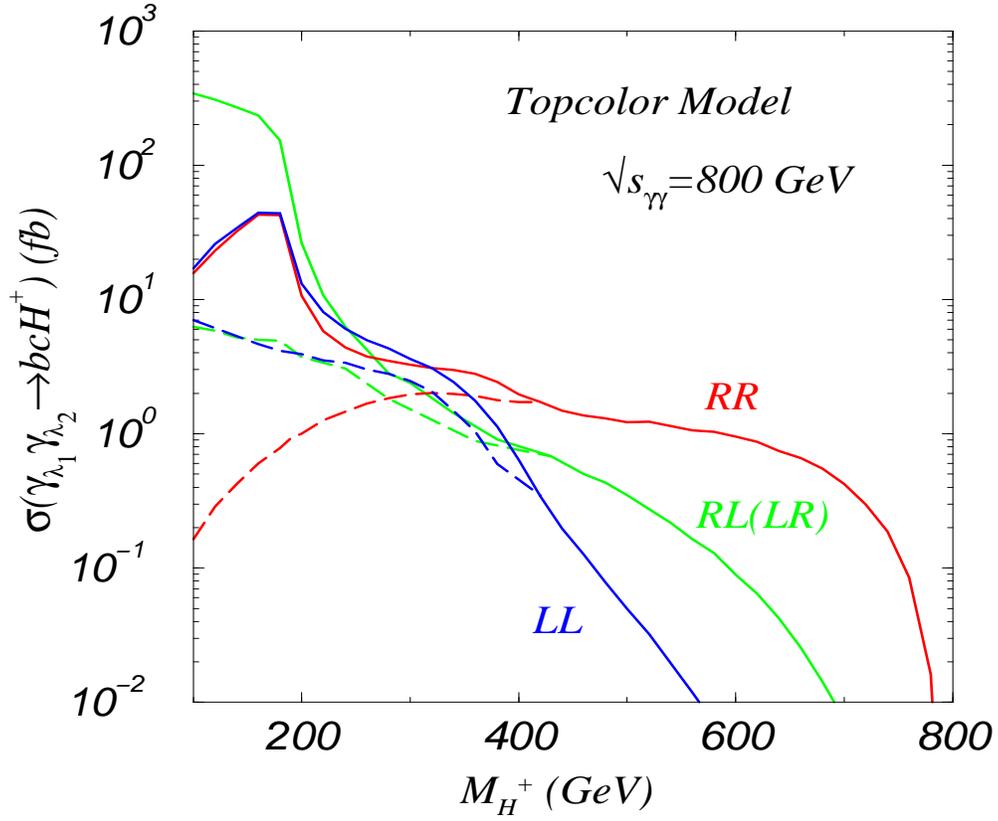}
\end{center}
\vspace*{-6mm}
\caption{Cross sections of
$\gamma_{\lambda_1}\gamma_{\lambda_2} \to b {\bar c} H^+$
         at $\sqrt{s_{\gamma \gamma}}=800$ GeV in
polarized photon collisions for the TopC model
[cf. Eq.\,(\ref{eq:bcH-TopC-def})].
Solid curves are the results without any kinematical cut, and
 dashed curves are the results with the kinematical cut
 specified in the text [cf. Eq.\,(\ref{eq:kin-cut})].
}
\label{fig:bch_topc_pol}
\end{figure}

\begin{figure}[H]
\begin{center}
\includegraphics[width=13cm,height=11cm]{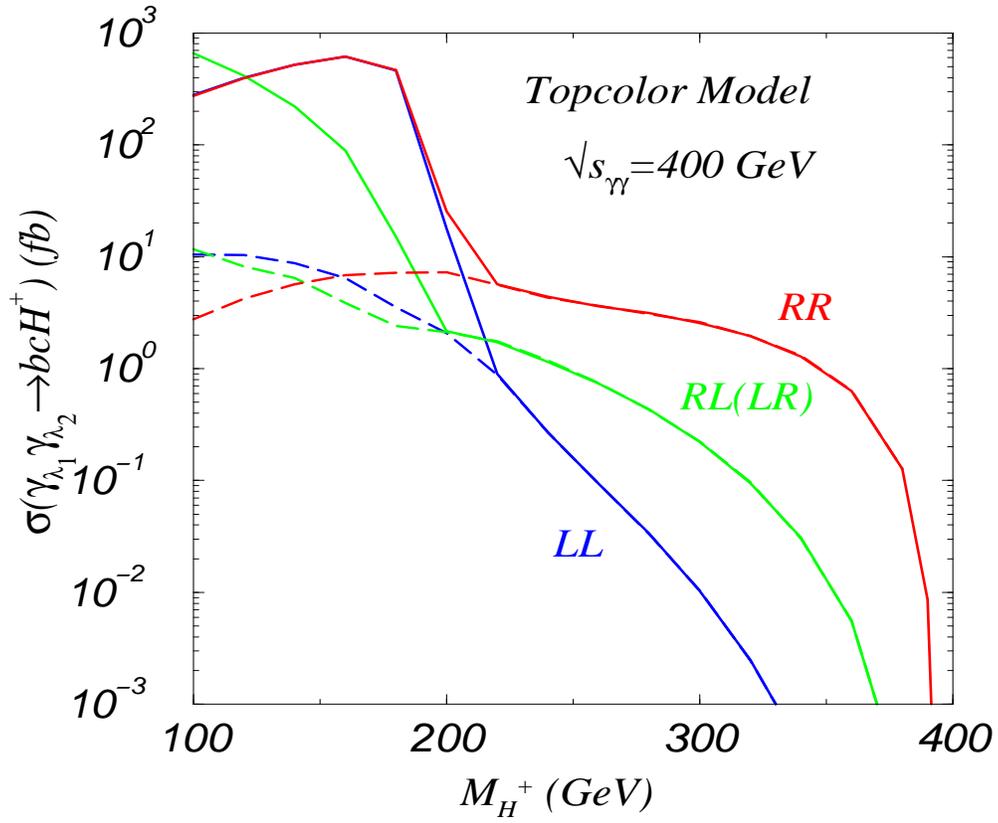}
\end{center}
\vspace*{-6mm}
\caption{Same as Fig.~\ref{fig:bch_topc_pol}, but for
$\sqrt{s_{\gamma \gamma}}=400$ GeV.
}
\label{fig:bch_topc_pol_2}
\end{figure}

\begin{figure}[H]
\begin{center}
\includegraphics[width=13cm,height=9.5cm]{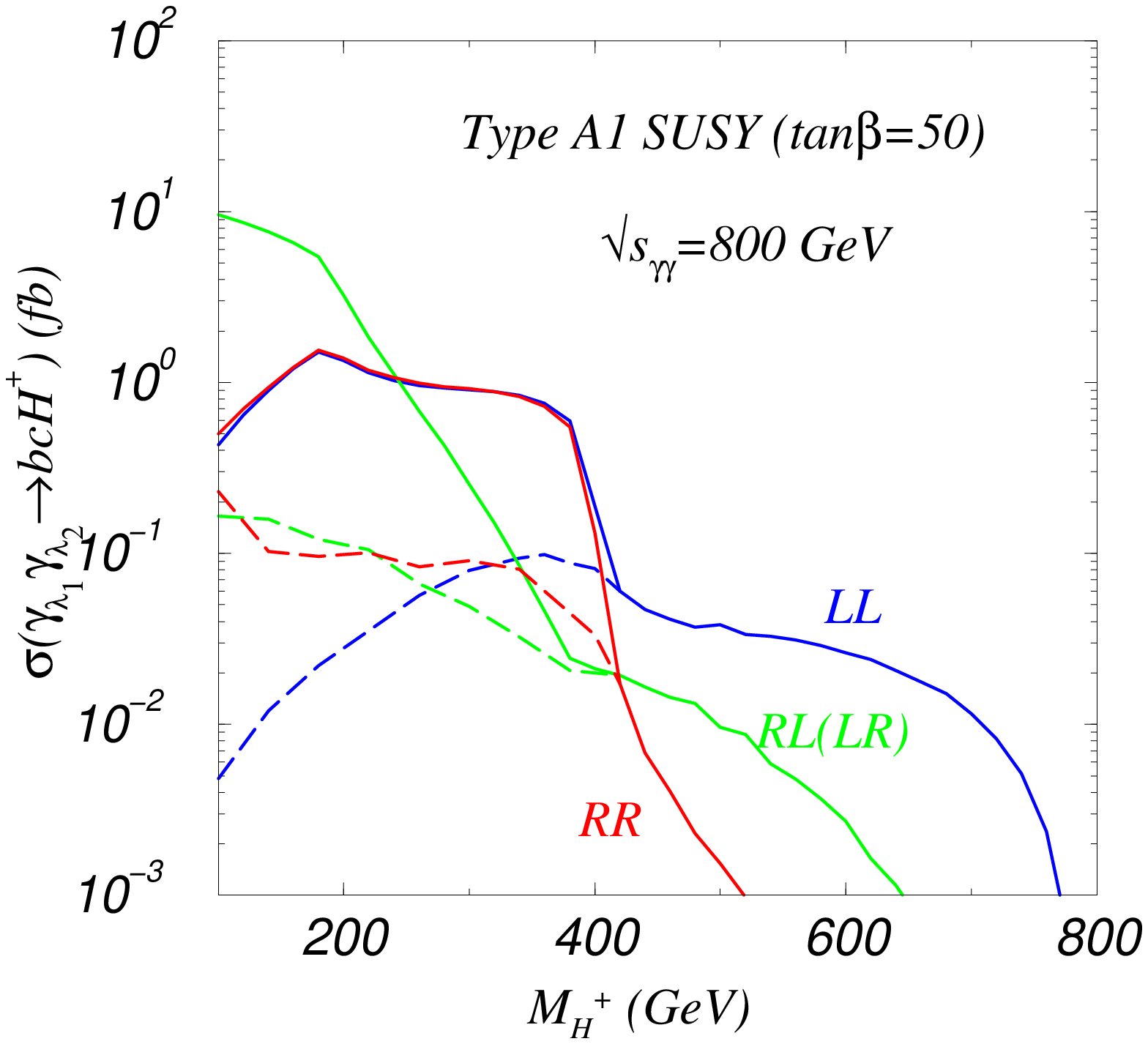}
\end{center}
\vspace*{-7mm}
\caption{Cross sections of
$\gamma_{\lambda_1}\gamma_{\lambda_2} \to b {\bar c} H^+$
         at $\sqrt{s_{\gamma \gamma}}=800$ GeV in
polarized photon collisions for the Type-A SUSY model
[cf. Eq.\,(\ref{eq:bcH-MSSM-def})].
Solid curves are the results without any kinematical cut, and
 dashed curves are the results with the kinematical cut
 specified in the text [cf. Eq.\,(\ref{eq:kin-cut})].
}
\label{fig:bch_mssm_pol}
\end{figure}

\begin{figure}
\begin{center}
\vspace*{-5mm}
\includegraphics[width=13cm,height=9.2cm]{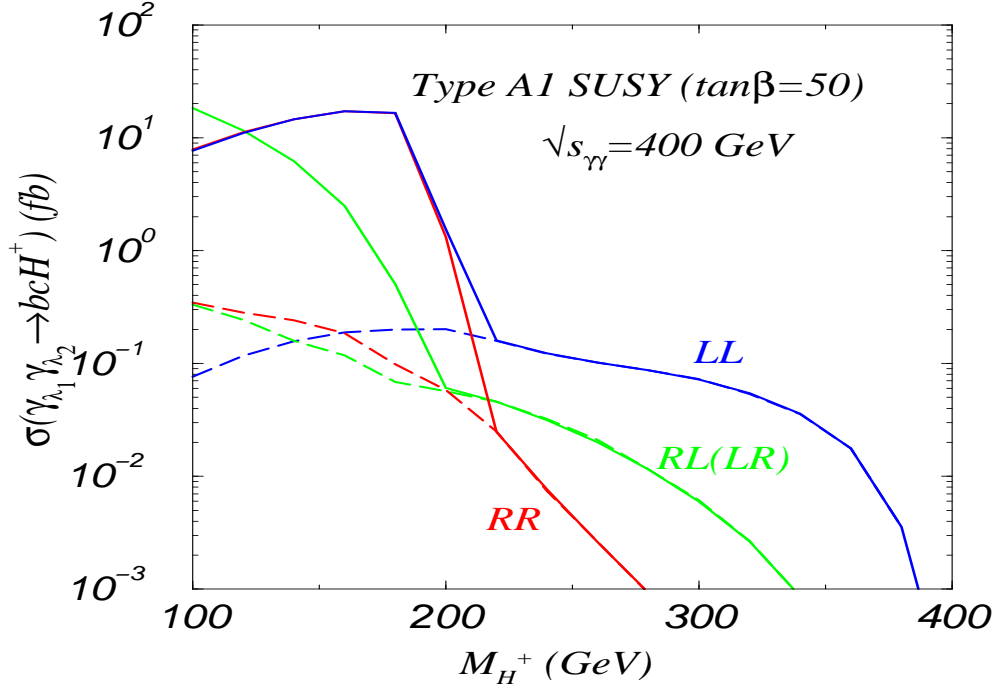}
\vspace*{-7mm}
\end{center}
\caption{Same as Fig.~\ref{fig:bch_mssm_pol}, but for
$\sqrt{s}=400$\,GeV.
}
\label{fig:bch_mssm_pol_2}
\end{figure}

From the above discussions we conclude that a photon-photon collider is
superior to an electron-positron collider for detecting a
heavy charged Higgs boson.
Moreover, a polarized photon collider can determine the
chirality structure of the fermion Yukawa couplings with the
charged Higgs boson via single charged Higgs production.
This point is illustrated as follows.
First, let us consider the case that $M_{H^\pm} > \sqrt{s}/2$.
As noted above, in this case, the production
cross section is dominated by the fusion
diagram $\gamma\gamma \to (c \bar c) (b \bar b)
\to  b \bar c H^+$. In the TopC model, because
$Y_L^{bc}=0$ (and $Y_R^{bc} \neq 0$), it corresponds to
$\gamma\gamma \to (c_R \ov{c_R}) (b_L \ov{b_L})
\to  b_L \ov{c_R} H^+$.
On the other hand, in the MSSM with stop-scharm mixings and
large $\tanb$, $Y_R^{bc} \sim 0$ (and $Y_L^{bc} \neq 0$),
it becomes
$\gamma\gamma \to (c_L \ov{c_L}) (b_R \ov{b_R})
\to  b_R \ov{c_L} H^+$.
Therefore, we expect that if both photon beams are
right-handedly polarized (i.e. $\gamma_R^{~} \gamma_R^{~}$), then
a TopC charged Higgs boson (i.e. top-pion) can be
copiously produced, while a MSSM charged Higgs boson
(with a large $\tanb$) is highly suppressed.
To detect a MSSM charged Higgs boson, both photon beams
have to be left-handedly polarized
(i.e. $\gamma_L^{~} \gamma_L^{~}$).
This is supported by an exact calculation whose results
are shown in Figs.\,\ref{fig:bch_topc_pol}
and \ref{fig:bch_topc_pol_2} for the TopC model
at two different collider energies.
A similar feature also holds for the MSSM after interchanging
the label of $RR$ and $LL$ in those figures,
which can be verified in Figs.\,\ref{fig:bch_mssm_pol} and
\ref{fig:bch_mssm_pol_2}.


In the following, we shall separately discuss the
feature of the polarized photon cross sections
for $M_{H^\pm}$ much less than $\sqrt{s}/2$ and for
$M_{H^\pm}$ slightly above $\sqrt{s}/2$.

The feature of the polarized photon cross sections
for $M_{H^\pm} < \sqrt{s}/2$ can be understood 
from examining the production process $\gamma\gamma \to H^+H^-$,
whose helicity amplitudes can be found in Eq.~(\ref{eq:pair}).
Let us denote
$\,\sigma_{ \lambda_{1} \lambda_{2}}^{\rm pair}$\, as the cross section
of \,$\gamma_{\lambda_1}^{~}\gamma_{\lambda_2}^{~}\to H^+H^-$\,.\,
We find that $\sigma_{LR}^{\rm pair}=\sigma_{RL}^{\rm pair}$, and they
dominate the total cross section when
\,$ M_{H^\pm}^2 \ll s$\,,\, while
$\sigma^{\rm pair}_{LL}$ and $\sigma_{RR}^{\rm pair}$
are equal and approach zero as \,$M_{H^\pm} \to 0$.\,
Since for \,$M_{H^\pm} < \sqrt{s}/2$\,  the bulk part of the
cross section of \,$\gamma\gamma \to b \bar c H^+$\,
comes from
\,$\sigma(\gamma\gamma\to H^+H^-)
   \times {\rm Br}(H^- \to b \bar c )$,\,
the $LL$ and $RR$ cross sections are smaller than the
$LR$ ($=RL$) cross sections as $M_{H^\pm}$
decreases, cf. Fig.\,\ref{fig:bch_topc_pol}.

As shown in  Fig.\,\ref{fig:bch_topc_pol}, the
polarized photon cross section $\sigma_{LL}$ is not zero
for $M_{H^\pm}$ slightly above $\sqrt{s}/2$, where the
on-shell $H^+H^-$ pair production channel is closed,
despite that the left-handed coupling
$Y_L^{bc}$ vanishes, cf. Eq.~(\ref{eq:bcH-TopC-def}),
in the TopC model.
This is due to the contribution from the diagrams in which
one of the charged Higgs boson is slightly off-shell
(as compared to its decay width), i.e.
from $\gamma \gamma \to H^+ H^{-*} (\to b {\bar c})$.
The similar argument also applies to the other models
but with different polarized states of the photon beams.

It is important to point out that the complete set of
Feynman diagrams have to be included to calculate
\,$\sigma(\gamma\gamma \to b \bar c H^+)$\, even when
\,$M_{H^\pm} < \sqrt{s}/2$\,  because of the requirement of
 gauge invariance.
To study the effect of the additional Feynman diagrams, other
than those contributing to the $H^+H^-$ pair production from
\,$\gamma\gamma\to H^+H^-(\to b \bar c )$,\,
one can examine the {\it single} charged Higgs boson cross section
in this regime with the requirement that the invariant mass of
\,$b \bar c$,\,  denoted as \,$M_{b\bar c}$,\, satisfies
the following condition\footnote{
These sample conditions are chosen to define the {\it single} charged
Higgs boson cross section, and they should be refined when a detailed
Monte Carlo simulation becomes available.
}:
\begin{eqnarray}
\label{eq:kin-cut}
|M_{b \bar c} & - & M_{H^\pm}|  ~>~   \Delta M_{b\bar c} \, ,
\qquad {\rm with}
\nonumber\\[3.5mm]
\Delta M_{b\bar c} & = &
\min \left[  25\,{\rm GeV},~
\max\left[ 1.18 M_{c\bar b} {\f{\,2 \delta m}{m}},\, \Gamma_{H^+} \right]
\right] \, ,
\nonumber\\[3mm]
\qquad \qquad
{{\,\delta m\,} \over m} & = & {0.5 \over \sqrt{M_{b\bar c}/2\,}\, }
\, ,
\end{eqnarray}
where $\,\dis\f{\,\delta m\,}{m}$\,   denotes
the mass resolution of the detector for observing the final state
$b$ and ${\bar c}$ jets originated from the decay of
\,$H^-$\,.\footnote{
Here, we assume the hadronic energy resolution for a jet with
energy $E$ (in GeV unit) is $50\% / \sqrt{E}$.
Moreover,
the full-width at half-maximum of a Gaussian distribution is
$1.18 \times 2 \,\sigma$, where $\sigma$ is
taken to be $M_{bc} \delta m /m$.
}
For instance, in Fig.\,\ref{fig:bch_topc_pol} the
set of dashed-lines are the polarized cross sections after
imposing the above kinematical cut.
With this cut, the total rate reduces by about one order of
magnitude for \,$M_{H^\pm} < \sqrt{s}/2$\,.\,
(However, this kinematical cut hardly changes
the event rate when \,$M_{H^\pm} > \sqrt{s}/2$\,.)
The effect of this kinematic cut on the
$RR$ and $LL$ rates are significantly different in the
low $M_{H^\pm}$ region. It implies that the $H^+H^-$ pair
production diagrams cannot be the whole production mechanism,
otherwise,
we would expect the rates of $RR$ and $LL$ be always equal
due to the parity invariance of the QED theory.
Again, a similar feature also holds for the MSSM after
interchanging the labels of $LL$ and $RR$.

Before closing this section, we remark that in the MSSM a heavy charged
Higgs boson $H^+$ can also be produced associated with a ${\bar c} s$
pair, whose production rate can be obtained by rescaling the cross
sections in Fig.\,\ref{fig:bch_mssm_tot} by the factor
$$
\(Y_{L(0)}^{sc}/Y_{L}^{bc}\)^2
~=~ 1.3 \, (\tanb)^2 \times 10^{-4}
$$
for \,$M_{H^\pm} > \sqrt{s}/2$\,.\,
Here, \,$Y_{L(0)}^{sc}=\dis\f{\sqrt{2}m_s}{v} \tanb\,$,\, and
the running mass of the strange quark at the scale of
$100$\,GeV is taken to be $m_s \simeq 0.1$\,GeV.\,
Hence, for $\tanb=30$, the production rate of
$\,scH^\pm$\, is down by a factor of 10\,, as compared
to the $b \bar c H^+$ rate with 
$Y_L^{bc}=0.05$, cf. Eq.~(\ref{eq:bcH-MSSM-def}).

\begin{figure}
\begin{center}
\includegraphics[width=13cm,height=8.5cm]{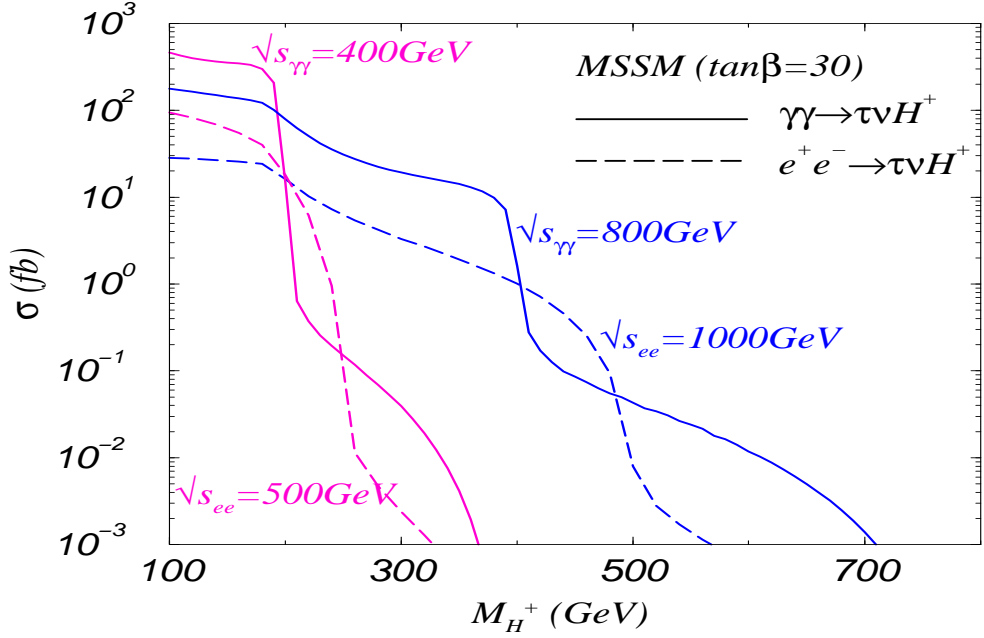}%
\end{center}
\vspace*{-7mm}
\caption{Cross sections of $\gamma\gamma\to\tau^- {\bar \nu} H^+$
         (solid curve)
         and $e^+e^-\to \tau^- \bar \nu H^+$ (dashed curve)
         for the MSSM [cf. Eq.\,(\ref{eq:taunuH-MSSM-def})]
         with unpolarized beams
         at $\sqrt{s_{\gamma \gamma}}=400$ GeV and $800$ GeV.
}
\label{fig:tnh_tot}
\end{figure}

\subsection{$\tau\nu H^\pm$ Production}

In the MSSM with a large $\tanb$ value, the
cross section of \,$\gamma\gamma\to\tau^- {\bar \nu} H^+$\,
can be quite sizable.
For the sample parameters chosen in Eq.\,(\ref{eq:taunuH-MSSM-def}),
its cross sections are shown in Fig.\,\ref{fig:tnh_tot} for
various linear colliders with unpolarized collider beams.
(Our results are consistent with the calculation in Refs.~\cite{KMO2,MO9}.)
Recall that we have chosen the sample parameters of the models so that
the Yukawa coupling of $\tau^-$-$\nu$-$H^+$ in the MSSM and
that of $b$-$c$-$H^+$ in the TopC model have the same magnitude
but opposite chiralities, as shown in Eqs.\,(\ref{eq:taunuH-MSSM-def})
and (\ref{eq:bcH-TopC-def}).
The gross feature of Fig.\,\ref{fig:tnh_tot} is similar to
Fig.\,\ref{fig:bch_topc_tot}.
However, a close examination reveals that the cross section
of $\gamma\gamma\to\tau^- {\bar \nu} H^+$ is smaller than that
of $\gamma\gamma \to b \bar c H^+$ at a fixed $M_{H^\pm}$
for $M_{H^\pm} > \sqrt{s}/2$.
For instance, for a 600\,GeV charged Higgs boson,
with its couplings given in Eqs.\,(\ref{eq:taunuH-MSSM-def})
and (\ref{eq:bcH-TopC-def}),
$\sigma(\gamma\gamma\to\tau^- {\bar \nu} H^+) \sim 0.01 \, {\rm fb}$
and
$\sigma(\gamma\gamma\to b \bar c H^+) \sim 0.3 \, {\rm fb}$, when
$\sqrt{s}=800$ GeV.
This difference can again be understood by examining the
Feynman diagrams.
In the scattering \,$\gamma\gamma\to b \bar c H^+$,\,
the total cross section is dominated by the fusion diagram
\,$\gamma\gamma \to (c \bar c) (b \bar b)
\to  b \bar c H^+$\, for  \,$M_{H^\pm} > \sqrt{s}/2$\,.\,
The contribution of this diagram is enhanced by
two collinear poles (in a $t$-channel diagram)
generated from  $\gamma \to c \bar c$ and
$\gamma \to b \bar b$ in high energy collisions.
However, in the scattering \,$\gamma\gamma\to\tau^- {\bar \nu} H^+$,\,
the dominant contribution in the large mass region comes from
the sub-diagram $\gamma \ov{\tau_R} \to H^+ \ov{\nu_L}$, and
contains only one collinear pole (in a $t$-channel diagram)
generated from  $\gamma \to \tau^- \tau^+$ in high energy collisions.
This is because photon does not couple to neutrinos.
Hence, the production rate of \,$\tau^-\bar\nu H^+$\,
is not as large as that of \,$b \bar c H^+$,\,
even when the relevant
Yukawa couplings are of the same magnitude
in both production channels.
For $M_{H^\pm}^{} < \sqrt{s}/2$ where $\gamma\gamma\to H^+H^-$
is kinematically allowed, the difference between
$\sigma(\gamma\gamma\to\tau^-\bar\nu H^+)$ and
$\sigma(\gamma\gamma\to b\bar c H^+)$ is caused
by the relative size of Br$(H^-\to \tau^-\bar\nu)$
and Br$(H^-\to b\bar c)$, cf. Fig.~\ref{fig:branch}.
\begin{figure}
\begin{center}
\includegraphics[width=14cm,height=8.5cm]{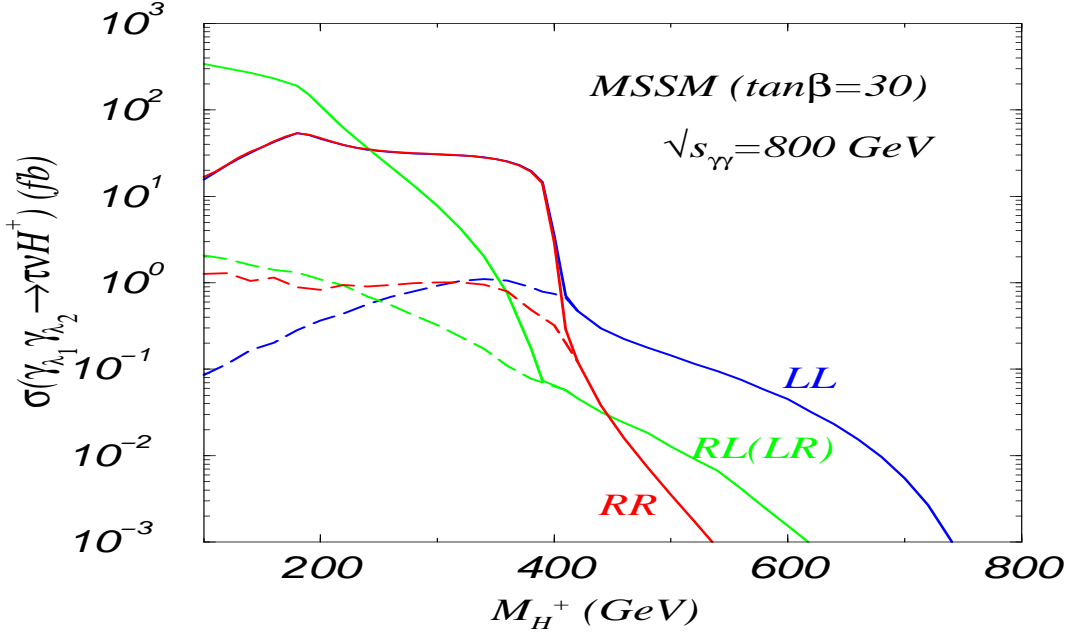}%
\end{center}
\vspace*{-5mm}
\caption{Cross sections of
$\gamma_{\lambda_1}\gamma_{\lambda_2} \to \tau^- {\bar \nu} H^+$
         at $\sqrt{s}=800$ GeV in
polarized photon collisions for the MSSM
[cf. Eq.\,(\ref{eq:taunuH-MSSM-def})].
Solid curves are the results without any kinematical cut, and
dashed curves are the results with the kinematical cut
 specified in the text [cf. Eq.\,(\ref{eq:kin-cut-tau})].
}
\label{fig:tnh_mssm_pol}
\end{figure}

\begin{figure}
\begin{center}
\includegraphics[width=14cm,height=10cm]{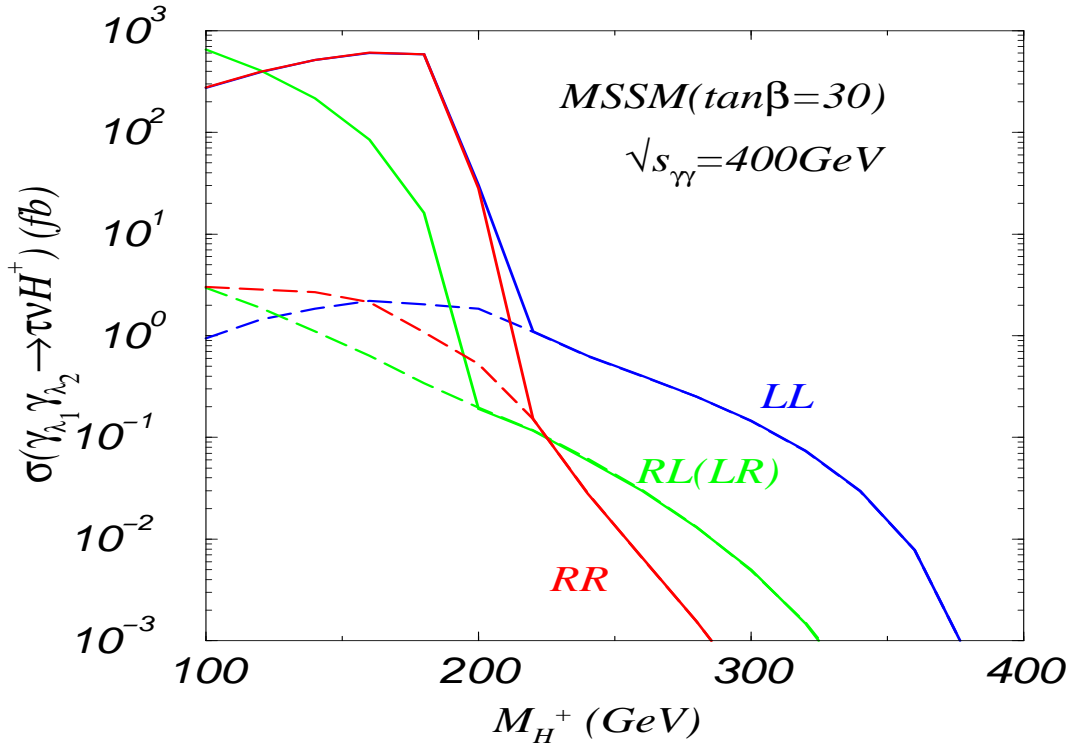}
\end{center}
\vspace*{-3mm}
\caption{Same as Fig.~\ref{fig:tnh_mssm_pol}, but for
$\sqrt{s}=400$ GeV.
}
\label{fig:tnh_mssm_pol_2}
\end{figure}

We also computed the production cross section
\,$\sigma(\gamma_{\lambda_1}^{} \gamma_{\lambda_2}^{} \to\tau^-\bar\nu H^+)$\,
in the polarized photon-photon collisions, and
the results are shown in Figs.\,\ref{fig:tnh_mssm_pol}
and \ref{fig:tnh_mssm_pol_2}.
As expected, the $LL$ rate is the dominant one when
$M_{H^\pm} > \sqrt{s}/2$\,,\, because the Yukawa couplings
\,$Y_R^{\tau\nu} = 0$\, and \,$Y_L^{\tau\nu} \neq 0$\,.\,
The {\it single} charged Higgs boson production rate
for \,$M_{H^\pm} < \sqrt{s}/2$\, is also calculated by
imposing the kinematical cut\footnote{
See footnotes 6 and 7, but for leptons. 
}:
\begin{eqnarray}
\label{eq:kin-cut-tau}
|M_{\tau \bar \nu} & - &  M_{H^\pm}|  ~>~  \Delta M_{\tau \bar \nu} \, ,
\qquad {\rm with}
\nonumber\\[2mm]
\Delta M_{\tau \bar \nu} & =  &
\min\left[ 25\,{\rm GeV},\,
\max\left[ 1.18 M_{\tau \bar \nu} \f{\,2 \delta m\,}{m},\,
\Gamma_{H^+} \right]
\right] \, ,
\nonumber\\[1mm]
\qquad \qquad
{\delta m \over m} & = & {0.5 \over ~\sqrt{M_{\tau \bar \nu}/2\,}~} \, ,
\end{eqnarray}
  and the result is shown in Figs.\,\ref{fig:tnh_mssm_pol}
and \ref{fig:tnh_mssm_pol_2}.
(In reality, $M_{\tau \bar \nu}$ should be replaced by, for instance,
 the transverse mass of the $\tau^- \bar \nu$ pair.)
For our choice of parameters in Eq.\,(\ref{eq:taunuH-MSSM-def}),
${\Gamma_{H^+}}$ is about
$0.54$\,GeV ($4.7$\,GeV) for a Higgs mass
$200$\,GeV ($400$\,GeV), and
correspondingly,
\,${\rm Br}(H^- \to \tau^- {\bar \nu})$\, is about
$0.69$ $(0.16)$.

\vspace*{5mm}
\section{Discussions and Conclusions}

In this work,
we have studied the single charged scalar production at polarized
photon colliders via the fusion processes
\,$\gamma \gamma \to b \bar c H^+$\, and
\,$\gamma \gamma \to \tau^- \bar\nu H^+$.\,
For the \,$b {\bar c} H^+$\, production,
we consider the flavor mixing couplings
of \,$b$-$c$-$H^\pm$\, generated
from the natural stop-scharm mixings in the MSSM, and
from the generic mixings of the right-handed top
and charm quarks in the dynamical Top-color model.
For the $\tau^- {\bar \nu} H^+$ production,
we consider the MSSM with a moderate to large $\tanb$.
We find that
the production rate of $H^+$ in the \,$\gamma\gamma$\, collisions
is much larger than that in the \,$e^-e^+$\, collision.
(Needless to say that the production rate of $H^-$ is the same as
$H^+$.)
Some of the results are shown in
Figs.\,\ref{fig:bch_topc_tot}, \ref{fig:bch_mssm_tot}
and \ref{fig:tnh_tot}.
For \,$M_{H^+}> \sqrt{s}/2$,\,
the cross section  of   \,$\gamma \gamma \to \tau^- {\bar \nu} H^+$\,
is smaller than that of \,$\gamma \gamma \to b {\bar c} H^+$\,
even when the corresponding Yukawa couplings are of
the same size.
This is because in high energy collisions there is only one
collinear pole
$\[\,\gamma \gamma \to (\tau^- \tau^+) \gamma
\to \tau^- {\bar \nu}  H^+\,\]$
in the scattering
\,$\gamma \gamma \to\tau^- {\bar \nu} H^+$,\,
but two collinear poles
$\[\,\gamma\gamma \to (c \bar c) (b \bar b) \to  b \bar c H^+\,\]$
in
\,$\gamma \gamma \to b \bar c H^+$.\,
The same reason also explains why in the large $M_{H^+}$
region the \,$e^+e^-$\, rate is smaller than
the \,$\gamma \gamma$\, rate by at least one
to two orders of magnitude,
since the \,$e^+e^-$\, processes contain only $s$-channel diagrams
and cannot generate any collinear enhancement factor to
the single charged Higgs boson production rate.
Furthermore, we show that it is possible to measure
the Yukawa couplings $Y_L$ and $Y_R$, separately,
at photon-photon colliders
by properly choosing the polarization states of the
incoming photon beams.
This unique feature of the photon colliders can
be used to discriminate new dynamics of the
flavor symmetry breaking.

To convert the cross sections, as shown in the above figures,
to the actual event rates,
one should take into account the corresponding collider
luminosity.
This is particularly important for calculating
the event rates at a photon collider, for the $\gamma \gamma$
luminosity depends on the energy of the photon beam which is
typically a distribution, in contrast to a fixed value, and the
degree of polarization of the initial state
photon will depend on its energy
\cite{telnov,kuhn}. Thus, the event rate at a photon collider
should be evaluated by convoluting the cross
section with
the $\gamma \gamma$ luminosity after accounting for the
energy dependence of the luminosity of the polarized photon 
beams.

\begin{figure}
\begin{center}
\includegraphics[width=11cm,height=8cm]{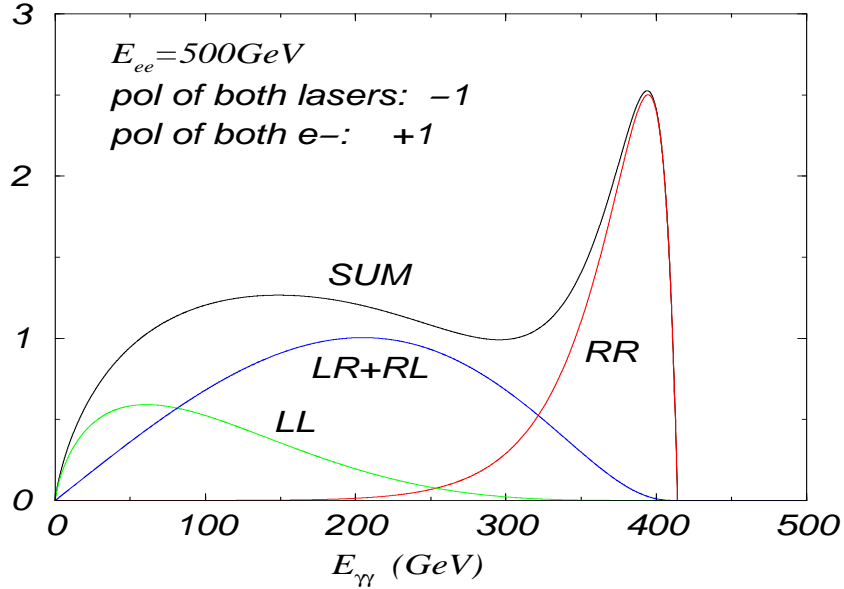}
\end{center}
\caption{  
 The luminosity of the photon beams $\gamma_{\lambda_1}
 \gamma_{\lambda_2}$ 
 produced from Compton back-scattering 
 as a function of the 
 c.m. energy for various polarization states of the two incoming photon
 beams, in the case that the laser beam is left-handedly polarized and the
 electron (or positron) beam is right-handedly polarized, and the 
 c.m. energy of the $e^-e^+$ collider is 500\,GeV.
 }
\label{fig:lumi}
\end{figure}

\begin{figure}
\begin{center}
\includegraphics[width=11cm,height=8cm]{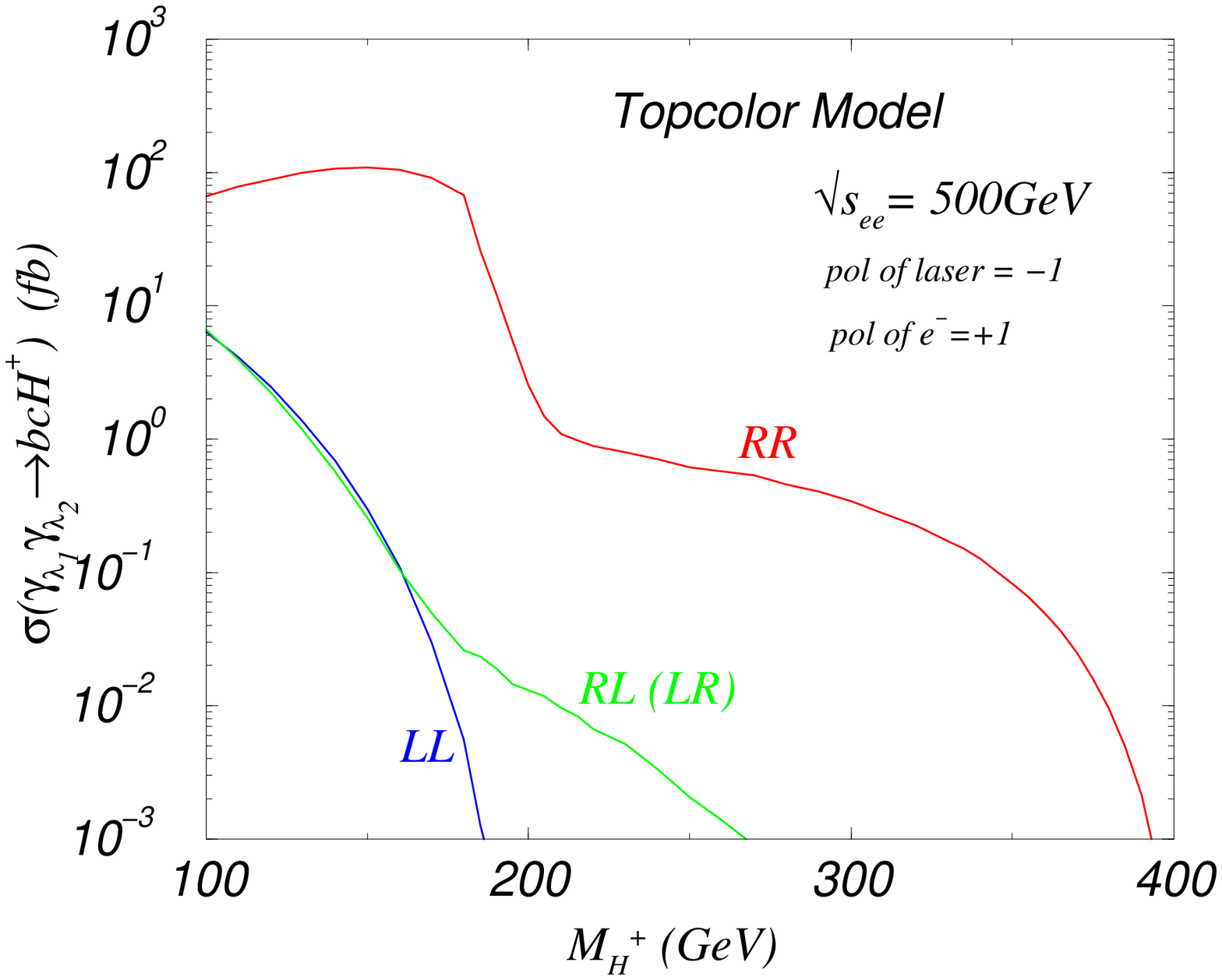}
\end{center}
\caption{  
 Cross sections (without any kinematic cut) for the TopC model
 after convoluting the constituent cross sections of 
 $ \gamma_{\lambda_1} \gamma_{\lambda_2} \to b {\bar c} H^+$, 
 cf. Fig.~\ref{fig:bch_topc_pol_2},
with the energy dependent  
 $ \gamma_{\lambda_1} \gamma_{\lambda_2}$ luminosity,
 cf. Fig.~\ref{fig:lumi}. 
 }
\label{fig:conv}
\end{figure}

To study the effect of the energy dependent luminosity of 
the polarized photon beam on the above analysis, we consider the model 
 suggested in Ref. \cite{pc} for producing a polarized photon beam from 
 the Compton back-scattering process ($e \gamma \to e \gamma$).
In this model, the $\gamma \gamma$ collider is based on a parent 
$e^-e^+$ (or $e^-e^-$) collider, and the luminosity distribution as a
function of the $\gamma \gamma$ c.m. energy is calculated by assuming
zero conversion distance for the $e^-$ (or $e^+$) beam.
As an example, let us consider the calculation that yields the result in 
Fig.~\ref{fig:bch_topc_pol_2}, but with convoluted $\gamma \gamma$
luminosities. 
 In the case that the laser beam is left-handedly polarized and the
 electron (or positron) beam is right-handedly polarized, the 
 luminosity of the photon beams produced from Compton back-scattering 
 as a function of the 
 c.m. energy for various polarization states of the two (recoiled) photon
 beams is depicted in Fig.~\ref{fig:lumi} based on the calculation  
 in Ref.~\cite{pc} with $x=4.82$, for a 500\,GeV $e^-e^+$ collider.
 Here, for simplicity, we have assumed a hundred percent polarized 
 $e^-$ (or $e^+$) beam.
 As shown in the figure, 
 the dominant (recoiled) photon polarization is the same as the electron 
 (positron)
 helicity, and the photon luminosity distribution peaks at high energy.
 When both the photon beams are right-handedly
 polarized (labelled as ``RR''), 
 the effective c.m. energy of the colliding photon beams is around
 400\,GeV for a 500\,GeV $e^-e^+$ collider. This justifies the
 approximation we made so far in our study. (The
 normalization of Fig.~\ref{fig:lumi}
  is such that the area covered by the curve labelled as ``SUM'', which
  is the result after summing up all the polarization states, is equal
  to the c.m. energy of $e^-e^+$, i.e 500 in this example.)
 After convoluting the constituent cross sections of 
 $ \gamma_{\lambda_1} \gamma_{\lambda_2} \to b {\bar c} H^+$, 
 cf. Fig.~\ref{fig:bch_topc_pol_2},
with the energy dependent  
 $ \gamma_{\lambda_1} \gamma_{\lambda_2}$ luminosity,
 cf. Fig.~\ref{fig:lumi}, we obtain 
 the result shown in Fig.~\ref{fig:conv}.
 Because we have chosen the polarization of the laser (and electron) 
 beam so that the 
 luminosity of the $\gamma_R \gamma_R$ state dominates in high energy
 region and the luminosity of 
 the $\gamma_L \gamma_L$ state is suppressed, hence,
  the difference between the $RR$ and the $LL$ rates 
 shown in Fig.~\ref{fig:conv} increases for a 
 larger $M_{H^+}$ as compared to  Fig.~\ref{fig:bch_topc_pol_2}.
 However, the magnitude of the ``RR'' cross section becomes smaller 
 because only some fraction of the
 produced photon beams is in the $\gamma_R \gamma_R$ state.
 For example, from Fig.~\ref{fig:bch_topc_pol_2}, the 
 cross section of 
  $ \gamma_{R} \gamma_{R} \to b {\bar c} H^+$ for $M_{H^+}=100$\,GeV 
  is 267\,fb when the c.m. energy of $\gamma \gamma$ is taken to be 
  400\,GeV.
For producing a 100\,GeV $H^+$, the effective integrated $\gamma_R
\gamma_R$ luminosity is about $1/3$ of the total $\gamma \gamma$
luminosity (i.e., after summing up all the polarization states of
$\gamma \gamma$), hence, the convoluted cross section can be estimated
to be 89\,fb ($=267\,{\rm fb}/3$).
This estimate agrees within a factor of 2 with the 
convoluted cross section  
exactly calculated in Fig.~\ref{fig:conv} 
which reads as 66\,fb for photons
produced from a 500\,GeV $e^-e^+$ collider using Compton back-scattering 
process.
Namely, the convoluted ``RR'' cross section is about $1/4$ of the
non-convoluted ``RR'' cross section.
The similar reduction factor for producing a heavier $H^+$ will be
somewhat bigger because the effective integrated $\gamma_R \gamma_R$
luminosity becomes smaller for $\gamma_R \gamma_R \to b {\bar c} H^+$.
For a 300\,GeV $H^+$, the convoluted ``RR'' cross section is about 
$1/7$ of the
non-convoluted ``RR'' cross section.

If we define the resolution power (${\cal A}$) of the 
polarized photon collider as 
\beq
{\cal A} \equiv { \sigma_{RR} - \sigma_{LL} \over \sigma_{RR} +
\sigma_{LL} }\, , 
\label{eq:power}
\eeq
then we conclude from the above discussion 
that a convoluted calculation predicts a stronger
resolution power of the polarized photon 
collider at the cost of a smaller cross section.\footnote{
A similar conclusion holds in the case that the 
$\gamma_L \gamma_L$ luminosity dominates, which can be generated 
by having the laser beam right-handedly polarized and the
 electron (or positron) beam left-handedly polarized.
}

According to the reports of the LC Working Groups in
Refs.\,\cite{TESLA} and \cite{JLC},
the integrated luminosity can reach about $500\,\ifb$
at a 500\,GeV LC, and $1000\,\ifb$ at an  1\,TeV LC.
Hence, we conclude that a polarized photon-photon collider is not only
useful for determining the \tx{CP} property of a neutral Higgs boson,
but also important for detecting a heavy charged Higgs boson
and determining the chirality structure of the
corresponding fermion Yukawa interactions with the
charged Higgs boson.

\vspace*{5mm}
\noindent
{\bf Acknowledgments}~~~\\
We thank Gordon L. Kane for valuable discussions
on the SUSY flavor mixings, and
Eri Asakawa and Stefano Moretti for comparing part of our results 
with their calculations.
This work was supported in part by the NSF grant PHY-0100677
and DOE grant DEFG0393ER40757.


\newpage
\begin{center}
{\bf {\Large References}}
\end{center}
\vspace*{-6mm}

\begin{thebibliography}{99}

\bibitem{Higgs}
P. W. Higgs, Phys. Lett. B{\bf 12}, 132 (1964);
             Phys. Rev. Lett. {\bf 13}, 508 (1964);
             Phys. Rev. {\bf 145}, 1156 (1966);
F. Englert and R. Brout,
             Phys. Rev. Lett. {\bf 13}, 321 (1964);
G. S. Guralnik, C. R. Hagen, and T. W. Kibble,
             Phys. Rev. Lett. {\bf 13}, 585 (1964).


\bibitem{SUSY}
P. Fayet and S. Ferrara,  Phys. Rept. {\bf 32}, 249 (1977);
H. P. Nilles,              Phys. Rept. {\bf 110}, 1 (1984);
H. E. Haber and G. L. Kane, Phys. Rept. {\bf 117}, 75 (1985);
and reviews in ``Perspectives on Supersymmetry'',
ed. G. L. Kane, World Scientific Publishing Co., 1998.


\bibitem{DSB}
For an updated comprehensive review of the dynamical symmetry
breaking and compositeness,
``Strong Dynamics and Electroweak Symmetry Breaking''
C. T. Hill and E. H. Simmons, 
{Phys. Rept. {\bf381}, 235 (2003),} \texttt{hep-ph/0203079}.



\bibitem{MSSM}
For recent reviews of MSSM, see, {\it e.g.,}
H. E. Haber, Nucl. Phys. Proc. Suppl. {\bf 101}, 217 (2001)
[\tx{hep-ph/0103095}] and \tx{hep-ph/0212136};  
G. L. Kane, {\it Lectures at the Latin American School,
SILAFAE III} [\tx{hep-ph/0008190}].


\bibitem{Hill}
C. T. Hill,
Phys. Lett. B{\bf 345}, 483 (1995) [\tx{hep-ph/9411426}]; and
Phys. Lett. B{\bf 266}, 419 (1991).


\bibitem{tbH}
F. M. Borzumati, in $e^+e^-$ Collisions at 500\,GeV, ed. P. M. Zerwas,
DESY 93-099, p.261-268, \tx{hep-ph/9310348};
J.~A. Coarasa, J.~Guasch, J.~Sola and W.~Hollik,
\Journal{\PLB}{442}{326}{1998} [\tx{hep-ph/9808278}];
J. A. Coarasa, D. Garcia, J. Guasch, R. A. Jimenez
and J. Sola, \Journal{\EPC}{2}{373}{1998} [\tx{hep-ph/9607485}].


\bibitem{gbHt}  J. Gunion, G. Ladinsky, C.-P. Yuan, {\it et al.,}
                in {\it Proceedings of
                 Snowmass Summer Study,} pp.\,59-81,
                Snowmass, Colorado, 1990 (World Scientific,
                Singapore, 1992);
                V. Barger, R. J. N. Phillips and D. P. Roy,
               \Journal{\PLB}{324}{236}{1994} [\tx{hep-ph/9311372}];
               F. Borzumati, J. Kneur, N. Polonsky,
               \Journal{\PRD}{60}{115011}{1999} [\tx{hep-ph/9905443}];
               A. Belyaev, D. Garcia, J. Guasch, J. Sol\`{a},
               JHEP {\bf 0206}, 059 (2002)
               [\tx{hep-ph/0203031}];
               T. Plehn, \Journal{\PRD}{67}{014018}{2003} [\tx{hep-ph/0206121}].



\bibitem{hy} H.-J. He and C.-P. Yuan, \Journal{\PRL}{83}{28}{1999}
             [\tx{hep-ph/9810367}].


\bibitem{bhy} C.\,Balazs, H.-J. He, C.-P. Yuan,
             \Journal{\PRD}{60}{114001}{1999}
             [\tx{hep-ph/9812263}].


\bibitem{dhy} J.L. Diaz-Cruz, H.-J. He, C.-P. Yuan,
              \Journal{\PLB}{530}{179}{2002}
              [\tx{hep-ph/0103178}].


\bibitem{xx}
For a recent application to the LHC-CMS analysis,
S. R. Slabospitsky, CMS-Note-2002/010,
[\tx{hep-ph/0203094}].


\bibitem{ppHW} A.A. Barrientos Bendezu and B. A. Kniehl,
               \Journal{\PRD}{59}{015009}{1999} [\tx{hep-ph/9807480}],
               \Journal{\PRD}{61}{097701}{2000} [\tx{hep-ph/9909502}],
               \Journal{\PRD}{63}{015009}{2001} [\tx{hep-ph/0007336}];
               O. Brein, W. Hollik and S. Kanemura,
               \Journal{\PRD}{63}{095001}{2001} [\tx{hep-ph/0008308}].


\bibitem{ppHW2}  S. Moretti and K. Odagiri,
                \Journal{\PRD}{59}{055008}{1999} [\tx{hep-ph/9809244}].


\bibitem{ppHpHm} A. A. Barrientos Bendezu and B. A. Kniehl,
               \Journal{\NPB}{568}{305}{2000} [\tx{hep-ph/9908385}];
                 O. Brein and  W. Hollik,
               \Journal{\EPC}{13}{175}{2000} [\tx{hep-ph/9908529}];
                 A.~Krause, T.~Plehn, M.~Spira and P.M.~Zerwas,
               \Journal{\NPB}{519}{85}{1998};
                 S.S.D.~Willenbrock, \Journal{\PRD}{35}{173}{1987}.

\bibitem{eeHpHm} S.~Komamiya,
                  \Journal{\PRD}{38}{2158}{1988};
                 A.~Djouadi, J.~Kalinowski and P.M.~Zerwas,
                  \Journal{\ZPC}{74}{569}{1993};
                 A.~Djouadi, J.~Kalinowski, P.~Ohmann and P.M.~Zerwas,
                   \Journal{\ZPC}{74}{93}{1997};
                 J.~Guasch, W.~Hollik, A.~Kraft,
                   \Journal{\NPB}{596}{66}{2001}.

\bibitem{eeHpHm2}
                 D.~Bowser-Chao, K.~Cheung, S.~Thomas,
                   \Journal{\PLB}{315}{399}{1993};
                 M.~Drees, R.M.~Godbole, M.~Nowakowski, S.D.~Rindani,
                   \Journal{\PRD}{50}{2335}{1994}.

\bibitem{eeWH}
S.-H. Zhu,  \tx{hep-ph/9901221};
S. Kanemura, \Journal{\EPC}{17}{473}{2000} [\tx{hep-ph/9911541}],
A. Arhrib, M. Capdequi Peyranere,
W. Hollik and G. Moultaka,
               \Journal{\NPB}{581}{34}{2000} [\tx{hep-ph/9912527}],
H. E. Logan, S. Su, \Journal{\PRD}{66}{035001}{2002} [\tx{hep-ph/0203270}].


\bibitem{KMO1} S.~Kanemura, S.~Moretti and K.~Odagiri,
               JHEP {\bf 0102}, 011 (2001) [\tx{hep-ph/0012020}],
               S.~Kanemura, S.~Moretti and K.~Odagiri,
               Proceedings of Linear Collider Workshop 2000 at
               Fermilab [\tx{hep-ph/0012020}].

\bibitem{hky} H.-J. He, S. Kanemura, C.-P. Yuan,
              \Journal{\PRL}{89}{101803}{2002}
              [\tx{hep-ph/0203090}].

\bibitem{pc} I.F.~Ginzburg, G.L.~Kotkin, S.L.~Panfil, V.G.~Zerbo
             and V.I.~Telnov, {\it Nuc. Inst. Methods} 5 (1984).

\bibitem{ecfa} ECFA/DESY Photon Collider Working Group,
             (B. Badelek et al.) [\tx{hep-ex/0108012}];
             American Linear Collider Working Group,
                                 [\tx{hep-ex/0106056}].

\bibitem{JLC} ACFA Linear Collider Physics Working Group,
              `` {\it Particle Physics Experiment at JLC} ''
              [\tx{hep-ph/0109166}].

\bibitem{CPprop1} B. Grzadkowski and J. F. Gunion,
                 \Journal{\PLB}{294}{361}{1992}
                 [\tx{hep-ph/9206262}].

\bibitem{CPprop3} M.~Kr\"amer, J.~K\"uhn, M.L. Stong,
                P.M.~Zerwas,
                \Journal{ZPC}{64}{21}{1994}.

\bibitem{CPprop2} E. Asakawa, S. Y. Choi, K. Hagiwara and J.-S. Lee,
                 \Journal{\PRD}{62}{115005}{2000}
                 [\tx{hep-ph/0005313}].

\bibitem{RevLC}
For reviews,
M. S. Chanowitz, Nucl. Instr. \& Meth.
A{\bf 355}, 42 (1995);
H. Murayama and M. E. Peskin, Ann. Rev. Nucl. Part. Sci.
{\bf 46}, 533 (1996) [\tx{hep-ex/9606003}];
E. Accomando {\it et al.,}
Phys. Rept. {\bf 299}, 1 (1998)
[\tx{hep-ph/9705542}].

\bibitem{lcwg} 
Mayda Velasco, et al., $\gamma \gamma$ working group, at Arlington Linear Collider Workshop,
Jan 9-11, 2003, the University of Texas at Arlington, TX, USA.  

\bibitem{telnov} V.~Telnov, {\it Nuc. Inst. Methods} {\bf A 294} 72 (1990);
                 I.~Ginzburg, G. Kotkin, V.~Serbo and V.~Telnov
                 {\it Nuc. Inst. Methods} {\bf A 205} 47 (1983),
                 {\it ibidem} {\bf A 219} 5 (1984).

\bibitem{ko} S.~Kanemura, K.~Odagiri, [\tx{hep-ph/0104179}].

\bibitem{kmo_eg} S.~Kanemura, S.~Moretti, K.~Odagiri,
              \Journal{\EPC}{22}{401}{2001}.

\bibitem{CCBVS}
J. A. Casas and S. Dimopoulos,
Phys. Lett. B{\bf 387}, 107 (1996) [\tx{hep-ph/9606237}].

\bibitem{FCNC}
M. Misiak, S. Pokorski, J. Rosiek,
``Supersymmetry and FCNC Effects'',
\tx{hep-ph/9703442},
in {\it Heavy Flavor II,}  pp.\,795,
eds., A. J. Buras and M. Lindner,
Advanced Series on Directions in High Energey Physics,
World Scientific Publishing Co., 1998,
and references therein.


\bibitem{QCDeff}
E.~Braaten, J.P.~Leveille, \Journal{\PRD}{22}{715}{1980};
M.~Drees, K.~Hikasa,       \Journal{\PRD}{41}{1547}{1990};
                           \Journal{\PLB}{240}{455}{1990},
                            (E) {\bf B 262}, 497 (1991);
A.~Djouadi, M.~Spira, P.M.~Zerwas, \Journal{\ZPC}{70}{427}{1996}.

\bibitem{hbb}
C.\,Balazs, J.L. Diaz-Cruz, H.-J. He, T.\,Tait, C.-P. Yuan,
\Journal{\PRD}{59}{055016}{1999} [\tx{hep-ph/9807349}].


\bibitem{efflag}
M. Carena, S. Mrenna, C.E.M. Wagner,
\Journal{\PRD}{60}{075010}{1999}
[\tx{hep-ph/9808312}].


\bibitem{dmb}
R. Hempfling, \Journal{\PRD}{49}{6168}{1994};
L. J. Hall, R. Rattazzi, U. Sarid,
\Journal{\PRD}{50}{7048}{1994}
[\tx{hep-ph/9306309}];
M.~Carena, M.~Olechowski, S.~Pokorski, C.E.M.~Wagner,
\Journal{\NPB}{426}{269}{1994}
[\tx{hep-ph/9402253}].



\bibitem{hwz} M. Capdequi Peylanere, H. E. Haber and P. Irulegui,
              \Journal{\PRD}{44}{191}{1991};
              A. Mendez and A. Pomarol,
              \Journal{\NPB}{349}{369}{1991};
              S. Kanemura,
              \Journal{\PRD}{61}{095001}{2000}
              [\tx{hep-ph/9710237}].


\bibitem{KMO2} S.~Moretti and S.~Kanemura, \Journal{\EPC}{29}{19}{2003}. 

\bibitem{MO9}
S.~Moretti,
talk at the International Workshop on Linear Colliders (LCWS2002),
Korea, August 26-30, 2002
[\tx{hep-ph/0209210}].

\bibitem{kuhn} J.~K\"uhn, E. Mirkes and J. Steegborn,
   \Journal{ZPC}{57}{615}{1993}

\bibitem{TESLA} ECFA/DESY Linear Collider Physics Working Group,
                TESLA Technical Design Report Part III:
                `` {\it Physics at an $e^+e^-$ Linear Collider} ''
                [\tx{hep-ph/0106315}].


\end{thebibliography}
\end{document}